# Constraining the Formation of the Four Terrestrial Planets in the Solar System

Patryk Sofia Lykawka[1] and Takashi Ito (伊藤孝士)[2]

[1] School of Interdisciplinary Social and Human Sciences, Kindai University, Shinkamikosaka 228-3, Higashiosaka, Osaka, 577-0813, Japan; patryksan@gmail.com

[2] National Astronomical Observatory of Japan, Osawa 2-21-1, Mitaka, Tokyo, 181-8588, Japan

## ABSTRACT

To reproduce the orbits and masses of the terrestrial planets (analogs) of the solar system, most studies scrutinize simulations for success as a batch. However, there is insufficient discussion in the literature on the likelihood of forming planet analogs simultaneously in the same system (analog system). To address this issue, we performed 540 N-body simulations of protoplanetary disks representative of typical models in the literature. We identified a total of 194 analog systems containing at least three analogs, but only 17 systems simultaneously contained analogs of the four terrestrial planets. From an analysis of our analog systems, we found that, compared to the real planets, truncated disks based on typical outcomes of the Grand Tack model produced analogs of Mercury and Mars that were too dynamically cold and located too close to the Venus and Earth analogs. Additionally, all the Mercury analogs were too massive, while most of the Mars analogs were more massive than Mars. Furthermore, the timing of the Moon-forming impact was too early in these systems, and the amount of additional mass accreted after the event was too great. Therefore, such truncated disks cannot explain the formation of the terrestrial planets. Our results suggest that forming the four terrestrial planets requires disks with the following properties: 1) Mass concentrated in narrow core regions between ~0.7–0.9 and ~1.0–1.2 au; 2) an inner region component starting at ~0.3–0.4 au; 3) a less massive component beginning at ~1.0–1.2 au; 4) embryos rather than planetesimals carrying most of the disk mass; and 5) Jupiter and Saturn placed on eccentric orbits.

# 1. INTRODUCTION

The formation of the four terrestrial planets in the inner solar system remains an unsolved problem in planetary sciences. Terrestrial planet formation is also linked to Moon formation, the origin of water and life on our planet, the orbital and compositional structure of the asteroid belt, the orbital architecture of the giant planets, and the properties of the protoplanetary disk that formed the planets (Morbidelli et al. 2012). Clearly, a successful model for the formation of the four terrestrial planets can provide important insights into such fundamental aspects of the solar system. In particular, we believe that *simultaneously* reproducing the orbits and masses of the four terrestrial planets is required for a successful model.

The terrestrial planets formed via the accretion of embryos and planetesimals embedded in the protoplanetary disk. The growth of disk planetesimals/embryos is an active area of research, for which many mechanisms have been proposed (e.g., cumulative accretion, Wetherill & Stewart 1989, Kokubo & Ida 2000, Goldreich et al. 2004; gravitational collapse, Morbidelli et al. 2009; pebble accretion, Levison et al. 2015; Chambers 2016). Although those models can yield diverse disk outcomes, a typical outcome is that of a protoplanetary disk consisting of several embryos and a population of small planetesimals. Previous studies have shown that under those conditions, a few terrestrial planets can form in the inner solar system (Chambers 2001; O'Brien et al. 2006; Raymond et al. 2009; Morishima et al. 2010; Lykawka & Ito 2013; Fischer & Ciesla 2014; Izidoro et al. 2015; Haghighipour & Winter 2016; Walsh & Levison 2016; Clement et al. 2018).

Several models of terrestrial planet formation have achieved a number of common successes such as producing massive planets akin to Venus and Earth and obtaining planets on stable orbits with relatively small eccentricities ($e$) and inclinations ($i$). However, most studies have difficulties in reproducing the dynamically cold (small $e/i$) orbits of Venus and Earth (Raymond et al. 2018). Other recent models have successfully produced planets with masses comparable to that of Mars after considering narrow protoplanetary disks[1] (e.g., disks with mass concentrated at semimajor axis, $a = 0.7$–$1.0$ au) instead of the more extended disks used in earlier models (Hansen 2009; Jacobson & Morbidelli 2014; Walsh & Levison 2016; Raymond & Izidoro 2017). Additionally, systems in which Jupiter and Saturn start with eccentricities comparable to or slightly higher than their current values appear to be a key feature to obtain the aforementioned successes (O'Brien et al. 2006; Raymond et al. 2009; Walsh & Morbidelli 2011; Haghighipour & Winter 2016). Indeed, instabilities experienced by the giant planets may yield such higher eccentricities and produce narrow disks from originally extended protoplanetary disks (Clement et al. 2018, 2019a). Narrow disks may also be the result of dynamical truncation by Jupiter owing to interactions with nebular gas in the solar system's infancy (Walsh et al. 2011). Thus, the disk properties and giant planet orbital evolution both play an important role in shaping the inner solar system.

Despite the aforementioned advances, reproducing the orbits and masses of the terrestrial planets (planet analogs) remains elusive. Indeed, forming the four terrestrial planets in the solar

---

[1] Kokubo & Ida (2000) were perhaps the first to use similarly narrow disks in terrestrial planet formation, but their goal/emphasis was on forming Earth.

system requires addressing the following issues.

**(1) Where are the Mercury analogs?** Studies on the formation of Mercury based on N-body simulations are scarce in the literature[2]. The reasons for this are twofold. To obtain accurate results near Mercury's orbit ($a = 0.39$ au) requires simulations of terrestrial planet formation that are more computationally expensive. That is, using time steps up to ~4.5 days for the Wisdom–Holman type symplectic integration. Additionally, reproducing a planet with only 5.5% the mass of Earth at $a \sim 0.4$ au may require investigating disks with mass depletion near that region (Chambers 2001; O'Brien et al. 2006). Similarly, other models suggest that the origin of Mercury is connected to dynamical events that affected the region at $a < 0.5$ au during solar system's infancy (e.g., Volk & Gladman 2015; Raymond et al. 2016). To the best of our knowledge, only Hansen (2009) and Lykawka & Ito (2017) (hereafter LI17) met the time step condition above, and only the latter work investigated the in situ formation of Mercury by modeling the inner region near 0.4 au. Hansen (2009) serendipitously obtained two Mercury analogs with $a \leq 0.5$ au and 0.025–0.2 Earth masses ($M_\oplus$) using a truncated disk with the total mass concentrated at $a = 0.7$–1.0 au (the disk core region). Their findings suggest that Mercury may be a primordial embryo that was scattered from the core region to its current orbit. LI17 obtained nine Mercury analogs with final $a \sim 0.3$ au and median $m \sim 0.2$ $M_\oplus$ in simulations of disks containing a mass-depleted inner region at $a = 0.2$–0.5 au and extended to 3.8 au. The findings of L17 suggest that Mercury may represent an initially small embryo that grew by accreting both in situ and more distant embryos/planetesimals. However, these results could not fully explain Mercury's orbit and mass. In addition, Mercury–Venus pairs typically formed closer to the Sun than the observed distances and their mutual orbital separation was not reproduced either. More recently, Clement et al. (2019c) obtained 16 Mercury analogs with $a < 0.5$ au and $m < 0.2$ $M_\oplus$ from 100 simulations[3] including collisional fragmentation based on Hansen's initial conditions. However, overall, their results could not explain the orbit and high core mass fraction of Mercury, or the Mercury–Venus orbital separation. In summary, given these difficulties, forming a low-mass Mercury and a well-separated high-mass Venus should be considered new key requirements for successful terrestrial planet formation.

**(2) Are Venus analogs obtained consistently in systems with Earth analogs?** There is insufficient discussion in the literature about the likelihood of obtaining analogs of the Venus–Earth pair instead of only Earth analogs. Brasser et al. (2016a) was the first study to investigate this issue by statistically comparing their obtained Venus- and Earth-like planets. Clement et al. (2018, 2019a) identified Venus and Earth analogs, but they focused on forming Mars. Here, we propose that in addition to forming analogs of the Venus–Earth pair, obtaining systems with Venus analogs less massive than the Earth analogs should also represent an important constraint. Indeed, reproducing the orbits and masses of the Venus–Earth pair may provide new insights into the properties of the protoplanetary disk that formed both planets (e.g., core region size, presence of mass-depleted inner or outer regions, fraction of mass in embryos to that in planetesimals).

**(3) What is the likelihood of obtaining Mercury and Mars analogs in systems with Venus–Earth**

---
[2] The formation of Mercury has also been addressed by other methods, such as SPH simulations of giant impacts (e.g., Benz et al. 2007; Asphaug & Reufer 2014; Chau et al. 2018).
[3] Note that the time step used in those simulations was 6 days, which may be too large to probe the region at $a < 0.5$ au.

**pair analogs?** Recent terrestrial planet formation studies managed to statistically produce low-mass Mars analogs. In particular, there are currently five main competing models. 1) *Grand Tack*: the protoplanetary disk is truncated at ~1 au by perturbations of an inward-then-outward gas-driven migrating Jupiter within the first 1–3 Myr of the solar system history, leaving a disk with mass concentrated within that distance after the disk gas dispersal (Walsh et al. 2011; Jacobson & Morbidelli 2014; Walsh & Levison 2016; Brasser et al. 2016a); 2) *Empty Asteroid Belt*: embryos and planetesimals in the protoplanetary disk formed concentrated within a narrow belt at ~0.7–1 au (Hansen 2009; Drazkowska et al. 2016; Raymond & Izidoro 2017; Ogihara et al. 2018); 3) *Early Instability*: the protoplanetary disk is perturbed by the giant planets' instability that occurred within ~10 Myr after the disk gas dispersal, strongly depleting the disk mass beyond ~1.3 au (Clement et al. 2018, 2019a); 4) *Pebble Accretion*: embryos and planetesimals form preferentially in the inner and outer regions of the protoplanetary disk, respectively, and the disk mass is concentrated within ~1.5 au (Levison et al. 2015; Chambers 2016); 5) *Sweeping Secular Resonance*: the growth of embryos and planetesimals is inhibited beyond ~1–1.5 au by perturbations of secular resonances that swept the disk during the disk gas dispersal (Bromley & Kenyon 2017). In all these models, Mars analogs may form as a result of the disk mass depletion or absence of mass beyond ~1–1.5 au. Additionally, from the standpoint of the initial conditions of the disk after gas dispersal, models 1 and 2 are very similar while models 3–5 probably also share similar properties. However, in the majority of those studies, it remains unclear whether these Mars analogs were obtained in systems that *also* contained Venus and Earth analogs. Moreover, even if they were, it is also unclear what fraction these three-planet analog systems would represent compared to all the systems obtained in those studies, because their results are often presented with all planets mixed (e.g., in plots of distance vs. mass). Furthermore, Mercury and Mars analogs in those studies are often defined as planets from the mixed population that satisfy an arbitrary distance range: e.g., <0.5 au and 1.2–2.0 au, respectively. However, this approach can lead to incomplete classifications in a given system (e.g., by failing to identify more massive Mars analogs at <1.2 au or failing to properly identify planet analogs). Indeed, it is difficult to discriminate Venus from Earth analogs, unambiguously identify Mercury/Mars analogs, and avoid misclassifications in past studies. If the five models described above are successful in reproducing a low-mass Mars, how can the best among them be discriminated without a proper system classification?

Here, we argue that a better understanding of terrestrial planet formation can be obtained by scrutinizing Venus, Earth and Mars (or Mercury) analogs *in the same system*. At a minimum, such an analysis should consider water delivery to the four planets, the geological evolution of Earth, the relative timing of giant impacts (e.g., the Moon-forming giant impact), and the orbital structure of the asteroid belt. A statistical analysis of such analog systems and the identified Mars analogs can also help us to critically evaluate/compare the competing models of the origin of Mars. In short, terrestrial planet analog systems will allow more accurate and comprehensive investigations of these questions and provide new insights as well. However, addressing these questions using solely Earth-like planets, or planets obtained in systems that form excessively massive Mars-like planets may lead to misleading conclusions. In line with this reasoning, Jacobson & Walsh (2015) analyzed the growth of Earth based on their Earth-like planets selected from systems that contained Venus-, Earth-, and Mars-like planets. More recently, Clement et al. (2018, 2019a) found that 9–58% of their successful systems contained Venus, Earth, and Mars analogs. In conclusion, forming planet analogs of Mercury, Venus, Earth, and Mars in the same system should be the ultimate goal of any terrestrial planet formation model.

In this work, we developed a classification algorithm to determine the Mercury, Venus, Earth, and Mars analogs in each obtained system. The algorithm first identities the Venus–Earth analog pair, which then establishes the region boundaries used to identify Mercury and/or Mars analogs consistently. The planetary mass ranges considered for analog candidates were 0.03–0.17 (Mercury), 0.4–2.0 (Venus and Earth), and 0.05–0.32 $M_\oplus$ (Mars). A planet is defined as any object with mass $m \geq 0.03\ M_\oplus$ (a minimum of 50% the mass of Mercury). Conversely, objects with $m < 0.03\ M_\oplus$ that started in the disk as embryos are defined dwarf planets. More details of our classification scheme are given in Section 2.1.

In order to address the main goals 1, 2, and 3, we performed extensive numerical simulations of terrestrial planet formation. In particular, we aim to reveal the likelihood of forming three- or four-planet analog systems and to constrain the protoplanetary disk initial conditions that favor their formation. We also aim to constrain the structure of the protoplanetary disk. For example, was the disk mass confined within a narrow core region at 0.7–1.0 au or 0.7–1.2 au? Did the disk consist solely of such a massive core region or did it also possess less massive components (e.g., an inner component near Mercury's orbit and/or an outer component beyond the core region)? In this work, we scrutinize a large suite of simulations with respect to a wide range of constraints and aim to better understand the conditions that gave rise to the inner solar system (tentative answers are discussed in Section 4).

**1.1 Main constraints on the inner solar system and measures of model success**

As discussed in the previous section, we believe that analog systems should serve as the baseline to verify the main constraints on the inner solar system. These constraints (in bold) and the associated model success criteria used in this work are summarized below:

**A. Formation of planets representative of Mercury, Venus, Earth, and Mars within ~2 au.** The orbits and masses of analogs representative of the terrestrial planets are identified by our classification algorithm (see Section 2.1 for details). For each final system, a minimum of three representative analogs are required to form. That is, an analog system must contain Venus, Earth, and at least a third analog (Mercury or Mars) in the same system and within 2 au.

**B. Terrestrial planet system properties.** The parameters of angular momentum deficit (AMD), radial mass concentration (RMC), and orbital spacing of adjacent planets (OS) are used to evaluate the orbital and mass distribution, and dynamical stability of the final systems (see Chambers 2001 for details about these parameters). The obtained systems should possess 50–200% of the RMC and OS, and a maximum of 200% of the AMD of the solar system (e.g., Jacobson & Morbidelli 2014).

**C. Absence of planets and existence of one dwarf planet in the asteroid belt.** Planets must not remain in the asteroid belt at $a \sim 2$–4 au for several hundred Myr after terrestrial planet formation (O'Brien et al. 2007). In contrast, a dwarf planet must exist in the asteroid belt to reproduce Ceres. We follow our systems for 400 Myr to verify the stability of such unwanted planets and the survival of dwarf planets.

**D. Timing of the giant impact that originated the Moon.** Early Earth probably experienced this giant impact ~30–150 Myr after the formation of the first solids in the solar system (Albarede 2009; Morbidelli et al. 2012; Jacobson & Morbidelli 2014). The starting conditions in our simulations translate this to a timing $t \sim 20$–140 Myr. In this work, based on studies of Moon-forming impacts (Jackson et al. 2018; Hosono et al. 2019), the collision of an impactor with at least 10% of the target's mass is considered a giant impact.

**E. Late veneer mass delivered to Earth.** The mass delivered to Earth via impacts of remnant embryo/planetesimals after the planet's last giant impact probably represents 0.1–1% of the Earth's mass (Raymond et al. 2014; Jacobson & Walsh 2015). However, there is no consensus about the definition of late veneer mass in the literature and its nature is still under debate; this constraint is not as well established as the other constraints in this section (Marty 2012; Brasser et al. 2016b; Rubie et al. 2016). Given these uncertainties, we relax the upper limit of the late veneer mass to 2%.

**F. Formation timescales of the four terrestrial planets.** Previous studies suggest that the Venus–Earth pair formed on timescales of 100 Myr (Kleine et al. 2009; Morbidelli et al. 2012; Raymond et al. 2014). The formation timescale of Mars is estimated to be a few Myr, but the uncertainties on such estimations are a factor of two or more, and thus a timescale of ~10 Myr is a reasonable assumption (Dauphas, N., Pourmand 2011; Mezger et al. 2013; Jacobson & Morbidelli 2014). The formation timescale of Mercury is not constrained yet. In this work, we focus on Mars'ss formation timescale constraint.

**G. Origin of water on Earth and the other terrestrial planets.** Water was likely delivered to Earth by impacts of objects that formed farther out than Earth in the protoplanetary disk (O'Brien et al. 2018). Estimates of the mass of water on Earth vary widely, such that Earth analogs are required to acquire water mass fractions (WMFs) between $2.5 \cdot 10^{-4}$ and $1 \cdot 10^{-2}$ (O'Brien et al. 2014; Izidoro et al. 2015; Jacobson & Walsh 2015; Raymond et al. 2018). Likewise, the same processes that brought water to Earth also operated during the formations of Mercury, Venus, and Mars. Mercury's WMF is unknown. Venus's WMF is poorly constrained, and thus we assume a WMF range of $5$–$50 \cdot 10^{-5}$ is needed for that planet[4]. Finally, we require Mars'ss WMF to be in the range $1$–$10 \cdot 10^{-4}$ (Lunine et al. 2003; Greenwood et al. 2018).

**H. Orbital structure of the asteroid belt.** The asteroids acquired their $a$–$e$–$i$ orbital distribution after the completion of both terrestrial planet formation and giant planet migration/instabilities (Izidoro et al. 2015; Clement et al. 2019b). However, the initial conditions of the asteroids in the early solar system are poorly constrained owing to the uncertainties in giant planet and planetesimal formation. For these reasons, we briefly discuss the asteroid belt in our simulations for completeness.

In this work, only analog systems were considered for the analysis against these constraints. However, there are two caveats that should be remembered when discussing the results.

---

[4] After Venus formation, the planet probably had a WMF on the order of Earth's ocean mass (~$10^{-4}$) (e.g., Morbidelli et al. 2000; Kulikov et al. 2006; Hashimoto et al. 2008; Greenwood et al. 2018). However, the uncertainties in those estimates are large. Here, we consider a Venusian WMF = $5$–$50 \cdot 10^{-5}$ to be an acceptable approximation.

Planetesimal-driven migration or dynamical instabilities of the giant planets may strongly perturb the planets and other bodies in the inner solar system (Brasser et al. 2009; Haghighipour & Winter 2016; Kaib & Chambers 2016; Clement et al. 2018). Indeed, obtaining a dynamically cold terrestrial planet system after those events is a subject of current research (Raymond et al. 2018). Additionally, similar to the results found in the literature, our results cover only a fraction of the solar system history, despite the fact that terrestrial planet systems may still be dynamically evolving after several hundred Myr. In summary, the migration/instabilities of the giant planets may affect the aforementioned constraints in some way, and the long-term dynamical evolution of a system may affect (e.g., Ito & Tanikawa 1999, 2002; Laskar 2008), in particular, constraints A, B, C, and H.

## 2. METHODS

We consider a primordial solar system of ~5–10 Myr of age consisting of the giant planets with their current masses and a protoplanetary disk containing several embryos and planetesimals. At this point, the disk gas mass is negligible compared to that of the solids (embryos, planetesimals; Haisch et al. 2001; Gorti et al. 2016), and thus gas dynamics are not considered here. We investigated the evolution of such primordial systems by performing 540 N-body numerical simulations, as described below.

We considered four initial orbital architectures for Jupiter and Saturn: 2:3 mutual mean motion resonance (MMR) low-$e$ (representative of the preferred Nice model; Morbidelli et al. 2007; Levison et al. 2011), 2:3 MMR moderate-$e$, 1:2 MMR high-$e$, and near-current orbits. We also tested four giant planets placed initially on mutual 2:3-2:3-3:4 MMRs (i.e., added Uranus and Neptune to the Nice model representative architecture). For brevity, these orbital configurations are referred to as JS23, JS23me, JS12he, NC, and JS23-4GPs, respectively (Table 1). These resonant orbital architectures cover the range of typical outcomes found by studies of planet–gas disk dynamics before giant planet migration/instabilities (Zhang & Zhou 2010; Pierens & Raymond 2011; D'Angelo & Marzari 2012; Pierens et al. 2014), and the NC configuration represents the giant planets after these events. In particular, these studies report several examples of Jupiter and Saturn evolving with eccentricities as high as represented by our JS23me and JS12he configurations, and thus it is important to consider their influence on terrestrial planet formation. To obtain the aforementioned resonant orbital architectures, similar to the methodology used in previous studies (e.g., Clement et al. 2018), we placed the giant planets outside the location of their mutual MMRs and forced them to migrate inward via fictitious forces until resonant capture. We then selected representative resonant systems that showed stability after evolving for 1 Gyr. In our simulations, to keep the model simple and to better explore the effects of disk properties, the giant planets do not migrate. The influence of giant planet evolution during terrestrial planet formation will be presented in our future publications[5].

---

[5] If Jupiter and Saturn evolved on eccentric orbits in the early solar system (as described by JS23me and JS12he), both planets may have acquired their current eccentricities via dynamical friction during planetesimal-driven planetary migration. Similarly, JS12he-like configurations might also produce Jupiter and Saturn with eccentricities slightly higher than their current values at the end of migration. If true, this would justify the initial conditions in the so-called EEJS

In modeling the protoplanetary disks, we divided the disk into three specific regions: inner, core, and outer. The core region concentrates most of the disk mass, and the inner and outer regions represent less massive components of the disk. The disk mass surface density increased in the inner region and decayed in the core and outer regions. In spite of that, all the disks described below started with ~2 $M_\oplus$ within 0.7–1.0(1.2) au. The disks typically contained several tens of embryos embedded in a sea of thousands of planetesimals (Table 2). These embryos were also individually tens to hundreds of times more massive than the planetesimals, which allowed an accurate representation of dynamical friction in the simulations (O'Brien et al. 2006). In particular, the dynamical friction was controlled by the ratio $r$, which represents the ratio of total disk mass distributed in embryos to that in planetesimals. Finally, the embryos were placed in the disk a few mutual Hill radii apart from each other. The procedures above followed typical methods in the literature (e.g., Chambers & Wetherill 1998; Chambers 2001; O'Brien et al. 2006; Raymond et al. 2009; Morbidelli et al. 2012; Raymond et al. 2014; Izidoro et al. 2015; Clement et al. 2018). The initial eccentricities and inclinations of the embryos and planetesimals were chosen randomly within the ranges for each disk model. The bulk density of the planetesimals and embryos was initially 3 g cm$^{-3}$.

We took into account the influence of the water ice line. This was modeled by considering a massive belt located in the outer regions of the disk, beyond 3.4 au (henceforth I-belts). The location of the ice line is currently at ~2.7 au, but it probably varied widely during planetesimal formation (Lunine 2006; Dodson-Robinson et al. 2009; Min et al. 2011). Indeed, the formation of planetesimals located near or beyond the ice line in the early solar system is an active area of research (Chambers 2016). In this way, we tested the hypothesis that icy bodies remain in relatively stable orbits within I-belts at the end of nebular gas dispersal. The disk mass enhancement in I-belts may have dynamically stirred the disk outer regions via embryo gravitational scattering, which could strongly affect Mars formation. I-belts may also offer new hints about the delivery of water to all the terrestrial planets.

We considered four models of protoplanetary disks: fiducial, truncated, depleted, and peaked.
*Fiducial disk*. This is the extended disk used in earlier studies in which the disk mass is distributed until the asteroid belt region. The disk surface density followed a power law with the exponent −1.5 across the entire disk. In disks that included an inner region at 0.3–0.7 au, the mass distribution in that region followed a linear increase in surface density (Chambers 2001; O'Brien et al. 2006; LI17). At the start of the simulations, the embryos had $e < 0.01$ and $i < 0.3$ deg, whereas the planetesimals had $e < 0.02$ and $i < 0.6$ deg. These slightly different $e/i$ distributions are consistent with oligarchic growth and gas disk studies (Kokubo & Ida 2000; Morishima et al. 2010). We tested a distribution of disk mass represented by $r = 1$. The individual embryo masses varied in the range of ~0.01–0.11 $M_\oplus$. Four Jupiter–Saturn orbital architectures were tested with fiducial disks: JS23, JS23me, JS12he, and NC (Table 1).
*Truncated disk*. This disk is used as representative of the typical truncated disks obtained in studies on the Grand Tack model (Walsh et al. 2011; Morbidelli et al. 2012; O'Brien et al. 2014; Jacobson & Morbidelli 2014; Brasser et al. 2016a; Walsh & Levison 2016). In that model, in order to explain the

---

model presented in Raymond et al. (2009).

current dichotomy of S-type and C-type asteroids (e.g., DeMeo & Carry 2014), S- and C-planetesimals are assumed to have formed in the cis- and trans-Jovian regions of the disk, respectively. After disk truncation, those studies found that the remaining embryos and S-planetesimals were concentrated between 0.7 and 1–1.2 au on low-$e$/-$i$ orbits and carried ~2 $M_\oplus$ of mass. In contrast, C-planetesimals acquired orbits covering a wide range of distances and eccentricities/inclinations and had a total mass of 0.1–0.3 $M_\oplus$. Here, we model truncated disks aiming to mimic the properties described above. At the start of the simulations, the embryos had $e < 0.05$ and $i < 1.5$ deg, whereas the S-planetesimals had $e < 0.15$ and $i < 5$ deg. The C-planetesimals were placed at $a > 1.0$ (or 1.2) au, $i < 20$ deg, and perihelion $q = 0.7$–1.7 au. We tested three distributions of disk mass, represented by $r = 1, 4,$ and 8. For a given $r$, the mass of the individual embryo was the same. The individual embryo masses varied within ~0.05–0.09 $M_\oplus$, with the values proportional to the $r$ value used. We also verified the influence of the core region size: 0.7–1.0 vs. 0.7–1.2 au. Two giant planet orbital architectures were tested with truncated disks: JS23 and JS23-4GPs.

*Depleted disk*. This disk is similar to that assumed in the Empty Asteroid Belt model. However, despite having the disk mass concentrated in the core region, the asteroid belt region would not be completely empty, but rather consist of a mass-depleted component. We modeled depleted disks using fiducial disks as the baseline (as modeled in Izidoro et al. 2014). The surface density in the depleted regions was found by multiplying the nominal surface density by a factor of 0.1 beyond 1–1.2 au. In disks that considered I-belts, the depleted region was defined between 1.0(1.2) and 3.4 au, whereas the I-belt was confined to $a = 3.4$–4.0(4.3) au such that the surface density was kept nominal within the belt. The embryos and planetesimals started the simulations with $e < 0.01$ and $i < 0.3$ deg, and $e < 0.02$ and $i < 0.6$ deg, respectively. We tested two disk mass distributions represented by $r = 1$ and 4. The embryos in the core (outer) region had individual masses within ~0.04–0.10 (0.003–0.018)[6] $M_\oplus$. The embryos inside the I-belts possessed individual masses of ~0.09–0.20 $M_\oplus$ (depending on the $r$ value used). In addition, we investigated the size of the core region: 0.7–1.0 vs. 0.7–1.2 au. Three Jupiter–Saturn orbital architectures were tested with depleted disks: JS23me, JS12he, and NC. For clarity, "depleted-IB" and "depleted-only" henceforth refer to depleted disks that did and did not include I-belts, respectively.

*Peaked disk*. This disk was inspired by the results of the pebble accretion model presented by Levison et al. (2015). In that model, at the end of gas dispersal, the embryos appear concentrated at ~0.7–2.7 au with individual masses decreasing with distance. The planetesimals follow a similar mass distribution and concentrate beyond ~1.5 au. Typically, a mass of 2.1–2.7 $M_\oplus$ is obtained inside 2 au and the embryos carry most of that mass. Pebble accretion models can yield various outcomes (Chambers 2016), such that the properties described above should not be considered unique. In an attempt to mimic these properties and their expected variation, we modeled six scenarios of peaked disks by testing two values of $r$ and three distinct mass distributions for the embryos. We also added an inner region to the disk to further explore the influence of that region. The distribution of mass in the disk resulted in a surface density that obeyed a power law increase from 0.4 to 0.9 au (exponents 1.5 for disks P9-20ir4 and P9-35ir4, and 2.5 for disk P9-20ir4#2) and a power law decay from 0.9 to 3.5 au (exponents −5.5, −6.5, and −4.25 for disks P9-20ir4, P9-20ir4#2, and P9-35ir4, respectively). The embryos were placed between 0.4 and 2.0 (3.5) au in disks P9-20ir4 and P9-20ir4#2 (P9-35ir4), and the planetesimals were placed beyond 0.9 au in all disks (Figure 1). At the start of the simulations, the embryos had $e < 0.001$ and $i < 0.03$ deg, whereas the planetesimals had $e < 0.01$ and $i < 0.3$ deg. The embryos had individual masses within ~0.0017–0.41 $M_\oplus$. Only the four-giant-planet orbital architecture JS23-4GPs was tested with peaked disks.

---

[6] The masses of the embryos in the outer region correspond to 20–110 times the mass of Ceres.

Previous terrestrial planet formation studies focused on the origin of Earth's water (e.g., O'Brien et al. 2018). Nevertheless, similar discussions are quite scarce for Mercury, Venus, and Mars in these studies. Thus, we consider water delivery to the four terrestrial planets in this work. For instance, these results may help us to better understand the presence of water ice and other volatile deposits on Mercury and Mars (Lunine et al. 2003; Peplowski et al. 2011; Eke et al. 2017) and constrain the water history on Venus (Hashimoto et al. 2008; Albarede 2009). The amount of water acquired by the planet analogs was estimated by considering the WMFs of the accreted embryos and planetesimals. The WMFs of these objects were set according to their initial locations in the disk: 0.001% at $a \leq 1.5$ au, 0.01% at $1.5 \leq a < 2$ au, 0.1% at $2 \leq a < 2.5$ au, 5% at $2.5 \leq a < 3$ au, and 10% at $a \geq 3$ au (e.g., Morbidelli et al. 2012). For disks with I-belts, the region with WMF = 20% was defined at $a > 3.4$ au. However, caution is needed with respect to water models. First, the initial WMFs are associated with the *current* radial structure in the asteroid belt, but that structure may have been different in the protoplanetary disk. Additionally, objects that formed beyond the ice line may have contained WMFs of tens of percent (e.g., Ciesla et al. 2015). For example, O'Brien et al. (2014) suggested that C-planetesimals in truncated disks may have contained WMF = 10%. In short, the distance ranges and WMFs adopted in this and past work are not unique (e.g., the often-adopted water model in Raymond et al. 2009). Here, we adopt the water model described above for simplicity and to facilitate the comparison with previous studies.

We considered a total of 54 simulation scenarios to cover the various parameters described above. In particular, 12, 12, 24, and 6 scenarios were considered to model the fiducial, truncated, depleted, and peaked disks, respectively. We tested each given scenario using 10 runs by varying the random seed numbers of the disks. We used a modified version of the MERCURY integrator, including real-time calculation of the bulk density/radii of embryos experiencing mass growth, general relativistic effects (which might be appropriate in the Mercury region at $a < 0.5$ au), and minor tweaks to perform all the simulations in this work (Chambers 1999; Hahn & Malhotra 2005; Kaib & Chambers 2016). In general, the densities of forming planets remained constant (i.e., 3 g cm$^{-3}$) in the simulations of previous terrestrial planet formation studies, despite the fact that the solar system terrestrial planets have larger values. In turn, the planets may have acquired artificially larger sizes and grown faster during those simulations, owing to the enhanced collisional cross sections. To prevent this from happening, we implemented a routine to update the bulk densities and radii of forming planets so as to approximately mimic those of the real planets following the method described in Brasser et al. (2016a). The giant planets and embryos were treated as massive bodies that perturbed each other, including mutual collisions. The planetesimals were also considered to be massive and to interact with the planets and embryos, but they did not mutually perturb each other.

Collisions were treated as perfectly inelastic, which is an acceptable assumption. Simulations including fragmentation led to quite similar system evolution and outcomes compared to those that did not include it (Kokubo & Genda 2010; Chambers 2013; Walsh & Levison 2016). One reason for this behavior is that fragments tend to be reaccreted by the forming planets. However, the production of fragments enhances the dynamical friction in the system, which can lead to less dynamically excited final planets (Chambers 2013; Clement et al. 2019a). Nevertheless, the large number of planetesimals in our simulations can mimic this dynamical friction, because a fraction of them can survive for tens of Myr. Overall, fragmentation is not crucial for the purposes of this work, but it will be considered in our future work.

All the simulations were evolved until 400 Myr. The time step was ~2.43 (1/150 yr), ~3.65 (1/100 yr), and ~4.57 (1/80 yr) days for disks that considered an inner disk edge at 0.3, 0.4, and 0.5 (or 0.7) au, respectively. These time steps are small enough to allow reliable calculation of planets with orbits similar to that of Mercury. Bodies that acquired heliocentric distances smaller than 0.1 au (0.15 au for time steps > 2.43 d) or larger than 20 au were eliminated from the simulations to avoid inaccurate orbital evolutions near the Sun and to save computer time for distant orbits, respectively. These bodies can collide with the Sun or can be gravitationally ejected by Jupiter in very short timescales, such that their influence in our results is negligible. Summaries of the initial conditions are given in Tables 1 and 2.

### 2.1 New classification algorithm for the inner solar system

First, the classification algorithm defines two regions in the inner solar system: the planetary region at $a \leq 2$ au and the asteroid belt region beyond 2 au. The choice of a threshold at 2 au is justified by two reasons. It is near both the current inner edge of the asteroid belt and the distance beyond which the gravitational scattering of a Mars-like planet becomes small. Planets must have $m \geq 0.03$ $M_\oplus$, and dwarf planets are defined as unaccreted embryos or grown embryos that satisfy $m < 0.03$ $M_\oplus$. For classification consideration, the system must contain at least three planets located within the planetary region, where at least two of them must have $0.4 < m \leq 2.0$ $M_\oplus$ to allow the determination of the Venus–Earth analog pair. If such conditions are not satisfied, the system is rejected, and no further classification takes place. The lower limit of 0.03 $M_\oplus$ comes from the assumption that a terrestrial planet should have at least 50% the mass of that of Mercury, and 0.4 and 2 $M_\oplus$ were chosen based on the assumptions that a Venus/Earth analog should have at least 50% of the mass of Venus and a maximum of 200% of the mass of Earth. Using the same mass criteria for Venus/Earth analogs allowed us to compare the mass distributions of Venus and Earth analogs without biases/artifacts that would arise if distinct criteria were considered. Additionally, even if Venus/Earth analogs were classified with $m > 0.5$ $M_\oplus$, this would not change the main results and conclusions of this study. The mass ranges for Mercury and Mars analogs are 0.03–0.17 and 0.05–0.32 $M_\oplus$, respectively. Those lower and upper limits are defined by assuming that the analogs should have half of and three times the mass of the real planets, respectively. Given the difficulty to reproduce the orbits and masses of the Mercury–Mars pair and to better understand the protoplanetary disk conditions that may reproduce them, the upper limits for both planets are slightly higher than those chosen for the Venus–Earth pair. It is worth noting that setting a Mars analog threshold at 1.25 or 2 times the mass of Mars does not dramatically change the main properties of the Mars analogs (Section 3.3.4) or the conclusions based on it.

Second, the algorithm initially assumes that the most and second most massive planets represent the Earth and Venus analogs of the system, respectively. In the case in which three or more planets with $m > 0.4$ $M_\oplus$ are present in the system, the algorithm proceeds as follows. If the second most massive planet is twice or more massive as the third, then the latter is discarded as a potential analog and the original assumption above remains unchanged. If not, only the most massive planet is kept as a potential Venus or Earth analog. Then, the algorithm must decide whether the second or the third most massive planet should be assigned as the second potential Venus or Earth analog. To do so, the algorithm first calculates the mass-weighted distances of the second and third most massive planets to that of the most massive one (indicated by subscript 1 below). The mass-weighted distance is given by $d_n = |a_n - a_1|(M_1/M_n)$, where $a$ is the average semimajor axis, $M$ is the mass, and the subscript $n$ indicates the $n$th most massive planet considered. If $d_3 < d_2$, the most massive and the third most massive planets are considered as the pair of Venus and Earth analogs. Instead, if $d_3 \geq d_2$, the two most massive planets are considered as the pair in question (the original assumption). Having determined the pair of analogs, the algorithm assigns the interior (exterior) planet among the two as the Venus (Earth) analog of the system. Finally, both analogs are automatically considered as the representative analogs of Venus and Earth, as any additional planet is classified later as planet-like by the algorithm.

Next, the algorithm defines the region interior to the Venus analog as the Mercury region, the region exterior to the Earth analog and with $a < 2$ au as the Mars region, and the region between the Venus and Earth analogs as the Venus–Earth region. In the Venus–Earth region, if the distance in Hill radii of a massive body with respect to the Venus analog is smaller than that to the Earth analog, the object is considered Venus-like, or Earth-like otherwise. Planets located in the Mars region that satisfy the Mars mass range constraint are classified as Mars analogs, while all other massive bodies in that region are classified as Mars-like objects. Conversely, the same procedure is performed in the Mercury region based on the Mercury mass range constraint. Irrespective of the number of identified Mercury (Mars) analogs, the representative analog is automatically considered the one located closer to the Venus (Earth) analog. Any dwarf planet found in the asteroid belt region is considered Ceres-like, and planets located in the same region are considered generic planets. Ultimately, for a given system, only one representative analog of each terrestrial planet is identified (if it exists), and additional analogs are only possible for Mercury or Mars. All the remaining planets are classified as planet-like, depending on their location in the planetary region.

In the last stage, if planet-like objects are found near the previously identified representative planet analogs, the classification algorithm disqualifies the latter by reclassifying them as planet-like in the following cases. If a large Mercury-like planet[7] formed in between Mercury and Venus analogs, the former analog is disqualified. If a planet with mass greater than half of that of the Venus analog formed adjacent to it, the analog is disqualified. The same applies to Earth analogs. If two or more planets, irrespective of their masses, formed between Venus and Earth analogs, both the analogs are disqualified. If a large Mars-like planet[7] formed between Earth and Mars analogs, the latter analog is disqualified. Reclassified systems with disqualified Venus/Earth analogs, or that possess only Venus and Earth as the remaining analogs, are excluded from the analysis of our results. In the end, only systems containing three or four representative analogs of real terrestrial planets are considered in the analysis. Note that planet-like objects, additional analogs (Mercury or Mars), or dwarf planets may be present in the system in addition to the representative planet analogs. Finally, the identified analog systems and their planets are evaluated against the inner solar system constraints in the next sections.

---

[7] In this work, large Mercury-like and Mars-like planets have masses $>0.17$ $M_\oplus$ and $>0.32$ $M_\oplus$, respectively.

## 3. MAIN RESULTS AND IMPLICATIONS

### 3.1 Terrestrial planet analog systems

From the combined results of 540 simulation runs, we obtained 21 Mercury–Venus–Earth, 156 Venus–Earth–Mars, and 17 Mercury–Venus–Earth–Mars terrestrial planet analog systems. Conversely, we identified 38, 194, 194, and 173 representative analogs of Mercury, Venus, Earth, and Mars, respectively, from these systems combined. All the additional Mercury/Mars analogs were taken into account when computing the system properties, but only the representative analogs were considered when discussing the properties of individual analogs (Section 3.3).

Of the four disk models considered in this work, fiducial disks produced the smallest fraction of analog systems (<8%), because most of the planets formed beyond the Earth analog were too massive to be considered acceptable Mars analogs. Truncated, depleted-only, depleted-IB, and peaked disks produced higher fractions of analog systems: 48%, 51%, 41%, and 28%, respectively. Peaked disks yielded a smaller fraction because they started with more massive feeding zones in their Mars regions, which reduced the likelihood of obtaining Mars analogs. These results are comparable to those found by Clement et al. (2019a). In particular, they found 58% and 13–36% analog system fractions for their annular disks based on Hansen (2009) and extended/annular disks perturbed by the giant planet instabilities, respectively.

Concerning truncated disks, those with $r = 4$ or 8 preferentially produced more analog systems. However, we found no similar trends for depleted disks, because the inclusion of an outer region masked the influence of the ratio $r$ (Section 3.2.1). Similarly, the giant planet orbital architecture also played a minor role in depleted disks (Section 3.2.2). Lastly, peaked disks with disk mass concentrated in narrow regions (P9-20ir4 and P9-20ir#2) produced more analog systems compared to a more extended mass distribution (P9-35ir4) (seven, seven, and three systems, respectively), because the Mars feeding zones were more massive in the latter disk setup. In summary, disks with an inner region component, truncated disks, or depleted disks are more likely to produce analog systems. More details are given in Table 3.

### 3.2 General trends and dependence on key parameters
#### 3.2.1 Ratio $r$

Low statistics prevented us from inferring the dependence on the ratio $r$ for peaked disks, and thus we summarize the results solely for truncated and depleted disks below. As discussed in Section 3.1, truncated disks modeled with $r = 4$ (or 8) rather than $r = 1$ yielded more successful results. In addition, disks with large $r$ also produced statistically fewer Mars analogs per system, yielded smaller late veneer mass fractions (for Earth analogs), and presented better mass fractions of Mercury/Venus and Mars/Earth analogs (Table 3). Furthermore, compared with disks with the initial $r = 1$, disks with $r = 4$ or 8 produced Mercury and Mars analogs that acquired orbits more distant from the Venus–Earth pair. Depleted disks with $r = 4$ rather than $r = 1$ also yielded better results according to the same system properties discussed above. Therefore, embryos rather than planetesimals probably dominated the protoplanetary disk mass after gas dispersal in the inner solar system.

These results can be understood according to the dynamical friction in the disk. The smaller the disk mass distributed in planetesimals (larger $r$), the weaker the dynamical friction (Jacobson & Morbidelli 2014). In fact, in systems with larger $r$, our embryos and forming planets acquired more dynamically excited orbits and experienced more mutual gravitational scattering events. Consequently, a higher number of less massive planets formed farther from the core region, which explains the higher fractions of Mercury or Mars analogs (hence, more analog systems as well) and the improved planetary mass fractions discussed earlier. Systems with weak dynamical friction also favor giant impacts on Earth to occur late during its formation. Thus, because the remaining disk mass was small at such late times, this also explains the smaller late veneer masses observed in our systems with larger $r$.

### 3.2.2 Giant planet orbital architecture

Fiducial disks that started with giant planets in the NC orbital architecture produced systems with a few Mercury and Mars analogs. However, the other giant planet architectures tested for this disk model did not produce promising results: the formation of Mercury- or Mars-like planets too massive to be considered analogs. Compared with the JS23, JS23me, and JS12he architectures, NC yielded more successful results because strong secular and mean motion resonances associated with Jupiter and Saturn led to more dynamical depletion of embryos/planetesimals beyond ~1.5 au (i.e., the higher the eccentricities of Jupiter and Saturn, the stronger those resonances; see e.g., Haghighipour & Winter 2016), thus increasing the probability of low-mass planets forming. Similarly, the JS12he architecture resulted in considerable dynamical excitation beyond ~2 au via the same resonances (Figure 2. See also Izidoro et al. 2016). For both the NC and JS12he architectures, embryos excited by such resonances also perturbed objects located in other regions of the disk (i.e., via secular conduction, as described by Levison & Agnor 2003). Because the region beyond ~1.5–2 au was initially mass depleted in depleted disks, the dynamical processes described above were less evident for those disks. Nevertheless, compared with NC and JS12he architectures, Jupiter and Saturn with smaller eccentricities as modeled by the JS23me architecture produced more systems that were too spatially compact, more Mars analogs per system, and relatively more massive Mars analogs in depleted disks (OS, Mpf, and $m_{Ma}$ quantities in Table 3). This was caused by the aforementioned weakened resonant and secular perturbations, which played a minor role in dynamically shaping the disk. In summary, even for disks with much less mass initially distributed beyond 1–1.2 au, Jupiter and Saturn in more eccentric orbits might help to produce less massive Mars analogs, lessen the chances of forming additional Mars analogs, and dynamically remove unwanted bodies in the asteroid belt.

### 3.2.3 Disk properties

How did the core region size (0.7–1.0 vs 0.7–1.2 au) affect the results? For truncated disks, T7-12 disks formed more planets than their T7-10 counterparts. This increased the chances of obtaining more analog systems and forming more additional Mars analogs within a given system. Indeed, two Mars analogs on spatially compact orbits are commonly seen in T7-12 disks. In such cases, the system mass is concentrated in five or more planets within 2 au, hence the large RMC and small OS obtained for these systems (Table 3). Unsurprisingly, these results reflect the initial conditions of larger core regions that facilitated the formation of spatially wider and more populated systems. Additionally, these differences became evident thanks to the dynamically cold environment of truncated disks, because the secular and resonant effects of the JS23 orbital architecture on the disk were negligible. However, these trends were not evident in depleted disks, because the latter considered an outer region embedded with embryos and distinct giant planet architectures that altogether perturbed the entire disk (Section 3.2.2). Apparently, the core region size played a minor role in those disks. With respect to peaked disks, considering the different initial total masses in the Mars region, P9-35ir4 produced almost no analogs of Mars, whereas the other two disk scenarios formed relatively more analogs. Therefore, the disk mass must decrease very steeply in the outer regions of peaked disks, as otherwise the Mars-like planets get too massive to be considered analogs.

Now, we discuss the influence of an inner region component in the disk. First, after comparing fiducial disks F7ir3 and F7, we found that three Mercury analogs were obtained in the former disk compared to one analog in the latter. In addition, when taking into account all the disks explored, the formation of less massive Mercury analogs (e.g., up to twice the real mass, $m < 0.11$ $M_\oplus$) occurred only in disks that considered inner regions, such as the fiducial disk F7ir3 and all the peaked disks (Figure 3). These results reflect the relatively small initial total masses in the Mercury region, which facilitated the in situ formation of less massive planets. Nevertheless, as discussed in Sections 3.3.1 and 3.4.2, the orbits of these Mercury analogs are dynamically too cold. In short, although the addition of an inner region increases the chances of producing more Mercury analogs of less mass, the details in that region (e.g., region boundaries, presence of planetesimals, embryos' masses, etc.) should affect the orbital properties of these analogs (Section 3.3.1).

We identified disk properties that affected the final location of Mercury and Venus analogs (Figure 3). Mercury analogs formed at medians $a = 0.52$–$0.53$ au for truncated and depleted disks, which is significantly farther from the Sun than Mercury. However, disks that included inner regions produced Mercury analogs with improved final locations, yielding median $a = 0.43$ au (Table 4). Venus analogs formed near the location of the disk inner edge for F5, F7, truncated, and depleted disks (Table 5). In particular, the truncated and depleted disks formed Venus analogs at a median $a = 0.65$ au, which is slightly closer to the Sun than Venus. However, the contribution of the inner region in disk F7ir3 caused their five Venus analogs to form at ~0.55–0.65 au. By comparing the medians *a* of our Venus analogs, we found that the size of the core region also influenced their final locations. Disk cores extended to 1.2 au (D/T7-12 disks) produced Venus analogs located 0.05 au farther from the Sun than the more compact disk cores extended to 1.0 au (D/T7-10 disks). For peaked disks, Venus analogs formed at smaller distances (median *a* = 0.66 au) for steeper disk mass distributions, as modeled in disk P9-20ir4#2, than for less steep distributions, as modeled in disk P9-20ir4 (median *a* = 0.77 au). Therefore, the location of the disk inner edge and the existence of an inner region component in the disk can play an important role in *simultaneously* determining the final locations of Mercury and Venus.

We also investigated the dependence of the final locations of Earth and Mars analogs on the disk properties (Figure 3). Although Earth analogs formed at final semimajor axes close to Earth's (Table 6), they were approximately 0.05–0.1 au closer to the Sun in T/D7-10 disks compared to their T/D7-12 counterparts. However, the ratio *r* also caused orbital spreading, such that the final locations of Earth analogs were slightly farther from the Sun for *r* = 4 and 8, owing to weaker dynamical friction in the disk (Section 3.2.1). For peaked disks, similar to Venus analogs, Earth analogs also formed at smaller distances (median *a* = 1.00 au) for steeper disk mass distributions than if less steep (median *a* = 1.13 au). Thus, the final locations of the Earth analogs resulted from details about the core region size and ratio *r*. However, the Mars analogs acquired a wide range of final orbits, such that the core region size apparently played a minor role for all the disk models considered. In contrast with the results for the Mercury analogs, we found no obvious disk property that could help to determine the final orbit of Mars analogs.

How did the inclusion of I-belts affect the results? Most of the main properties of analog systems obtained in depleted-only disks were similar to those obtained in depleted-IB disks. However, there were a few differences. First, massive icy embryos perturbed objects in the outer regions of the disk, causing them to be dynamically excited or removed from the system. Second, because of such additional perturbations, depleted-IB disks produced ~19% fewer representative Mars analogs (Table 7). Finally, a fraction of such massive embryos also survived in the asteroid belt region in 15/49 ~30% of the depleted-IB analog systems.

### 3.2.4 Other parameters

When considering subsets of truncated and depleted disks sharing the same initial *r* and Jupiter–Saturn resonance (2:3 MMR), the latter disks produced terrestrial planets on more dynamically excited orbits, slightly more massive Earth analogs compared to Venus analogs, and a fraction of Earth analogs that could acquire sufficient water (Table 3). These results were caused by the presence of outer regions that supplied additional mass and water to the Earth analogs. Compared to the JS23 orbital architecture, the more eccentric JS23me used in depleted disks also enhanced that mass supply and caused the extra excitation in the final planets. Concerning the long-term dynamical evolution of systems that considered the JS23-4GPs architecture, no such systems experienced a dynamical instability in truncated disks. Additionally, the results obtained for systems with only the Jupiter–Saturn pair and those with the four giant planets appear indistinguishable (a similar result was found by Walsh et al. 2011). On the contrary, three systems in peaked P9-35ir4 disks experienced a giant planet instability at late times (~300–400 Myr), thus representing 15% of the modeled 20 systems for this disk scenario in particular.

### 3.3 Properties of terrestrial planet analogs

We obtained the main properties of the representative analogs of Mercury, Venus, Earth, and Mars as identified in our analog systems (Section 3.1).

#### 3.3.1 Mercury analogs

Figure 3 shows that most representative Mercury analogs cannot reproduce the orbit and mass of Mercury (see also Table 4). Most of the Mercury analogs started as "seed embryos" within the disk core regions, slowly accreted mass and then moved inward to their final orbits. However, there are five exceptions. First, the single low-mass analog obtained in a depleted disk, which originated at ~2.1 au and that was later strongly scattered inward. Second, the four low-mass analogs that formed in their inner regions of fiducial and peaked disks. Notably, only seed embryos that started within ~0.4–0.7 au were able to acquire final $a \sim 0.4$ au and $m < 0.110$ $M_\oplus$ (twice the mass of Mercury). Finally, only three Mercury analogs acquired orbits and masses similar to those of Mercury: 1) $a = 0.36$ au, $e = 0.18$, $i = 6.7$ deg, $m = 0.053$ $M_\oplus$; 2) $a = 0.42$ au, $e = 0.19$, $i = 5.9$ deg, $m = 0.071$ $M_\oplus$, and; 3) $a = 0.44$ au, $e = 0.14$, $i = 4.2$ deg, $m = 0.054$ $M_\oplus$. In summary, with a few exceptions, our Mercury analogs are in general too massive and too dynamically cold.

Concerning the feeding zones of our Mercury analogs, they accreted mass from the core regions in all disks that did not include inner region components. For disks that included such components, the Mercury analogs acquired ~30–50% and ~50–90% of their masses from the inner regions in the F7ir3 (0.3–0.7 au) and peaked disks (0.4–0.9 au), respectively. In addition, the contribution of accreted embryos and planetesimals in the formation of our analogs varied significantly among the disk models. In truncated and depleted disks, the Mercury analogs started as massive seed embryos (i.e., with large fractions of the final mass) that later acquired their remaining mass via planetesimal accretion (~40% of the final mass). In contrast, the Mercury analogs started as small seed embryos in fiducial and peaked disks. Later, these analogs accreted their bulk mass by accreting mostly planetesimals (embryos) in fiducial (peaked) disks. As discussed in Section 3.2.3, these differences simply reflect the distinct initial conditions of our disk models. The Mercury analogs also acquired non-negligible fractions of their masses from the disk outer regions. In particular, these contributions were on average 3, 13–15, and approximately 10% for truncated, depleted, and peaked disks, respectively. The accretion of these materials also typically occurred at $t > 50$ Myr. Thus, Mercury analogs acquired water and possibly other volatiles during the late formation of these planets.

The disk models also predict different collisional histories for the Mercury analogs. All 16 analogs produced in truncated disks did not experience a single giant impact. In contrast, three out of eight and five out of six Mercury analogs experienced at least one giant impact in depleted-only and depleted-IB disks, respectively. Finally, the Mercury analogs obtained in fiducial and peaked disks experienced three or four giant impacts during their evolutions (Table 4).

### 3.3.2 Venus analogs

Our identified representative Venus analogs were highly successful in reproducing Venus's orbit and mass (Figure 3 and Table 5). In general, the Venus analogs started as seed embryos at ~0.7–0.9 au within the disk core regions before acquiring their final orbits.

Although the Venus analogs acquired their bulk masses from the core region in all the disk models, the contribution of the inner region component was on average ~30–40% in fiducial and peaked disks. Most of the Venus analogs started as embryos with <10% of the final masses. Later, these analogs accreted the remaining mass through rapid accretion of embryos and planetesimals. However, the Venus analogs obtained in peaked disks experienced a different accretion history. In particular, these analogs started as embryos with a median mass of 0.24 $M_\oplus$ (~30% the final mass), which later acquired their bulk masses by accreting mostly embryos. This result reflects the initial conditions in which embryos were more abundant in the inner/core regions of peaked disks. Lastly, the outer regions in truncated, depleted, and peaked disks contributed an average of 2, 7–8, and 10% of the Venus analogs' final mass, respectively. Therefore, the Venus analogs acquired sufficient water that may explain the estimated WMF for the real planet (footnote 4).

Finally, the Venus analogs experienced 3–8 giant impacts through their accretion histories. Additionally, the last giant impacts computed for all the analogs typically occurred very early, within ~3–8 Myr. In contrast, such impacts occurred late in fiducial disks (median 26 Myr).

### 3.3.3 Earth analogs

The main properties of our representative Earth analogs are summarized in Figure 3 and Table 6. It is worth noting that although the Earth analogs share several similarities with their Venus counterparts, a significant fraction of the Earth analogs are less massive than our planet. Nevertheless, this low-mass problem is not seen in fiducial and peaked disks. Overall, the Earth analogs started as seed embryos within the disk core regions at ~0.8–1.0 au before obtaining their final orbits.

The Earth analogs acquired their bulk masses from the core region in all disk models. Because our Earth analogs formed at distances greater than their Venus counterparts, the contribution of the disk inner components was not greater than ~10% in fiducial and peaked disks. The accretion history of the Earth analogs was very similar to that found for the Venus analogs, as discussed in the previous section. Similarly, our Earth analogs also acquired water from the accretion of outer region objects. In particular, these contributions were on average 2, 9–10, and 18% of the final planet masses in the truncated, depleted, and peaked disks, respectively. However, the obtained WMFs were in general below the minimum needed to explain Earth's bulk water content (Table 6).

Concerning giant impacts, the Earth analogs experienced 3–9 such events during their formation. The last giant impacts occurred after median 6–36 Myr, depending on the model used. Note that except in the fiducial disks, the medians are smaller than the minimum of 20 Myr that is needed to satisfy the Moon-forming giant impact. In particular, the systematic rapid formation of Earth analogs represents an outstanding problem for disks with mass solely concentrated at 0.7–1.0 au, which are the standard conditions after gas dispersal in the Grand Tack and Empty Asteroid Belt

### 3.3.4 Mars analogs

Overall, our identified 173 Mars analogs acquired median $a \sim 1.46$ au, $e \sim 0.04$, $i \sim 2.5$ deg, and $m \sim 0.20$ $M_\oplus$ (Figure 3 and Table 7). Despite the large number of analogs obtained, it remains challenging to reproduce the low mass of Mars and its moderately excited orbit. In particular, out of the 173 representative Mars analogs, 100 acquired masses less than two Mars masses (<0.214 $M_\oplus$) and only 45 reached 1.25 times the mass of Mars (<0.134 $M_\oplus$). Additionally, the Mars analogs obtained in truncated and peaked disks are dynamically too cold. The great majority of our Mars analogs started as seed embryos in the outer parts of the core regions in both truncated and depleted disks. Only five Mars analogs formed in fiducial disks originated within ~1.1–2.3 au, but four of them acquired final orbits more excited than that of Mars. Similarly, the 17 Mars analogs obtained in peaked disks started in the ~0.6–1.4 au distance range. Finally, overall, there was no obvious trend between the seed embryo initial location and the final location among all 173 Mars analogs.

In all disk models, the Mars analogs started as small seed embryos with tenths of the final masses. In particular, these fractions were approximately 40, 50 (25), 45 (30), and 55% for fiducial, truncated $r = 4$ or 8 ($r = 1$), depleted $r = 4$ ($r = 1$), and peaked disks, respectively. Later, the analogs acquired significant fractions of the remaining mass by accreting embryos/planetesimals from the core region. The outer regions also contributed to the analogs' final masses by 3% and 18% on average in the truncated and depleted disks, respectively. Unsurprisingly, this reflects the initial conditions of depleted disks that contained more extended outer regions. Additionally, the remaining mass of the Mars analogs in peaked disks was distributed as ~10, 30, and 5% for the inner (0.4–0.9 au), core (0.9–2.0 au), and outer regions (2.0–3.5 au), respectively. Finally, given the contribution of water-rich bodies located beyond ~2 au, the Mars analogs possibly acquired enough water to explain that inferred for the real planet (Table 7).

Overall, irrespective of the disk model, the Mars analogs experienced one to two giant impacts through their accretion histories. These giant impacts occurred after a median of 4–32 Myr, depending on the model used. It is worth noting that the systematic slow formation of the Mars analogs in fiducial, depleted-IB, and peaked disks is inconsistent with the inferred rapid formation of the planet (Section 1.1).

### 3.4 Four-terrestrial-planet analog systems (4-P systems)

In this section, we discuss the global properties of the obtained 17 Mercury–Venus–Earth–Mars analog systems and the details of the representative planet analogs formed in those systems. The level of success of the 4-P systems is discussed based on eight constraints C1–C8, defined as follows. System AMD = 0–0.0036 (C1), RMC = 44.9–179.4 (C2), OS = 18.9–75.4 (C3), and no planets in the asteroid belt (C4). Additionally, concerning the Earth analog, the last giant impact timing = 20–140 Myr (C5), late veneer mass fraction = 0.1–2% (C6), and WMF = $2.5 \cdot 10^{-4} – 1 \cdot 10^{-2}$ (C7). Finally, the Mars analog formation time is < 10 Myr (C8) (see Section 1.1 and Table 3 for details). The results discussed below are in broad agreement with the findings and mechanisms discussed in Sections 3.1–

3.3.

### 3.4.1 System properties

As measured by the system AMD, RMC, and OS quantities, 4-P systems are in general dynamically cold, highly mass concentrated, and too spatially compact (Figure 4). This means that constraints C1 and C2 are highly successful, but constraint C3 is not. In particular, our 4-P systems formed Mercury analogs too close to Venus analogs and in most cases, additional Mars analogs within ~1.5 au, which explains the difficulty in satisfying the C3 constraint. Additionally, we see no planets with $m > 0.05$ $M_\oplus$ left in the asteroid belt after 400 Myr of evolution. Indeed, given the initially low mass in the outer regions of depleted and peaked disks, only Ceres-like dwarf planets could survive in systems obtained from those disks. Thus, constraint C4 can be satisfied easily. Considering the high disk surface densities in truncated/depleted disks, approximately 80% (14 out of 17) of the Earth analogs formed rapidly enough to experience their last giant impacts in less than 20 Myr, which led to late mass accretions on the order of 10% the Earth's final mass. For these reasons, our 4-P systems have remarkable difficulty in satisfying constraints C5 and C6. Furthermore, it is clear that our 17 Earth analogs are too dry to explain Earth's water budget for the water model adopted in this work. Hence, constraint C7 is rarely satisfied. Finally, with a few exceptions, in general, the best Mars analogs acquired their bulk mass in less than 10 Myr, such that the associated constraint C8 is often satisfied.

Concerning the disk model used, 11, 5, and 1 of the 17 4-P systems were obtained for truncated, depleted, and peaked disks, respectively (Table A1). Disks with initial mass highly concentrated as modeled in the first two disk models appear to facilitate the formation of 4-P systems. Additionally, 12 out of the 17 systems were obtained in disks with ratio $r = 4$ or 8. Note that except for 4-P systems #15 and #16, the other systems have giant planets in their pre-migration orbital structure. Thus, it is possible that planets and dwarf planets are located beyond ~1.5 au may become dynamically unstable after the orbital rearrangements of Jupiter and Saturn (e.g., Clement et al. 2018).

### 3.4.2 Notes on the individual planet analogs in 4-P systems

Overall, the properties of the 17 representative planet analogs discussed below are similar to those found for all the analogs combined, as discussed in Sections 3.3.1–3.3.4 and summarized in Tables 4–7. Thus, in this section we provide only complementary details to deepen our understanding of each terrestrial planet (see also Figure A1).

First, the Mercury analogs formed farther from the Sun than Mercury and closer to Venus than in reality (Tables 4-5 and Figure 4). All analogs evolved in the core region during the first ~10–20 Myr and acquired their final orbits and masses by the end of 100 Myr of evolution. The feeding zones of the Mercury analogs suggest a significant in situ contribution to their formation, but the accretion of outer region objects was a common outcome. These analogs also acquired orbits that were too dynamically cold in truncated disks, suggesting that more dynamically active disk environments (such as modeled in depleted and peaked disks) may do a better job of explaining Mercury's excited orbit. Finally, only 3 out of the 17 Mercury analogs experienced at least one giant impact during their formation (systems #12, #15, and #17).

Concerning the Venus and Earth analogs, although they originated in the disk core region, their feeding zones were slightly shifted inward and outward, respectively. It is worth noting that the deviation was pronounced in the peaked disks. During the first 10 Myr of evolution, the analogs of both planets experienced 5–6 giant impacts, but three analogs of Earth experienced late giant impacts: 22, 38, and 53 Myr in systems #7, #14, and #11, respectively. Finally, after ~100 Myr of orbital/accretional evolution, the analogs of Venus and Earth acquired similar final distances compared to their initial ones.

Most of the Mars analogs (12 out of 17) formed closer to the Sun than Mars and closer to Earth than in reality (Tables 6–7 and Figure 4). Sixteen Mars analogs originated within the disk core region, evolved at 0.6–1.5 au, and acquired their bulk masses in less than ~10 Myr (the exception was in system #14, where the analog took ~100 Myr). Later, the Mars analogs evolved to their more distant final orbits during the remaining 390 Myr of evolution. During this time interval, these analogs also acquired a further small fraction of their final masses, including contributions of objects that were initially located beyond ~1.5 au. In particular, the analogs of Mars that acquired orbits/masses similar to those of Mars formed only in more dynamically active disks, such as truncated/depleted disks with $r = 4$ or 8, peaked disks, or giant planets on an NC or JS12he orbital configuration. Concerning giant impacts, whereas 12 of the Mars analogs experienced at least one such event during their accretion histories, 5 of them did not in systems #2, #4, #10, #11, and #15.

Note that important issues remain. The mass ratios of the Mercury–Venus and Mars–Earth analogs are systematically larger than those observed in the solar system. Additionally, a number of systems produced Venus analogs more massive than their Earth counterparts. Lastly, only six representative Mars analogs acquired $m \sim 0.1\ M_\oplus$, as found in systems #2, #4, #10, #11, #14, and #15.

## 4. DISCUSSION

We now discuss the ability of our 194 terrestrial planet analog systems to satisfy the main constraints of the inner solar system (Section 1.1).

*A. Formation of analogs of Mercury, Venus, Earth, and Mars (in terms of orbit and mass)* – Truncated, depleted, and peaked disks produced analog systems in several simulation runs. The initial high disk mass concentration near the region now occupied by Venus and Earth is the requirement to satisfy this constraint. However, the results did not allow us to unambiguously constrain whether this core region was confined at 0.7–1.0 or 0.7–1.2 au. More importantly, our results showed that: 1) only Venus and Earth are reasonably reproduced, but their mutual orbital separation and mass ratios are problematic; 2) it is difficult to obtain low-mass Mars analogs on moderately excited orbits; 3) it is rare to obtain Mercury analogs, irrespective of their masses or orbits; 4) Mercury and Mars analogs systematically form too close to Venus and Earth, respectively. These analogs also acquire excessively cold orbits. Intriguingly, the inclusion of collisional fragmentation does not help to reproduce either Mercury's peculiar orbit (Clement et al. 2019c) or Mars's low mass (Clement et al.

2018). Therefore, a more detailed exploration of initial disk conditions is needed to better reproduce the four terrestrial planets consistently.

*B. System properties* – A large fraction of systems produced in truncated and depleted disks yielded acceptable AMD, RMC, and OS values. In particular, the results indicate that protoplanetary disks with highly concentrated mass distributions are needed to satisfy this constraint. For this reason, steeper mass distributions would be needed to mimic that concentration in peaked disks. Additionally, the correlations of smaller RMCs/higher AMDs with larger $r$ found in truncated disks are in agreement with the findings of Jacobson & Morbidelli (2014). This strengthens the case that our truncated disks represented the typical outcomes in the Grand Tack model fairly well. Our results also indicate that forming a low-AMD system with the presence of Mercury and Mars distant from their massive neighbor planets remains a remarkable challenge.

*C. Absence of planets beyond 2 au and the existence of Ceres* – The majority of systems formed in truncated and depleted-only disks had no planets beyond 2 au, within the asteroid belt. With respect to Ceres, depleted and peaked disks produced systems that contained objects that could resemble the dwarf planet (e.g., Figure 4). The orbits and masses of these objects were similar to the initial conditions of embryos in the outer regions of those disks. Therefore, Ceres might be a leftover object from that population. Indeed, it is possible that additional planets or an excess population of dwarf planets were dynamically removed during giant planet migration/instabilities.

*D. Earth's last giant impact timing* – Except for the fiducial disks, the other disk models produced Earth analogs that experienced the last giant impacts within ~10 Myr, well before the conservative minimum 20 Myr required by giant impact models of the Moon. Indeed, these results suggest that it is difficult to conciliate this constraint with the rapid accretion generally expected in narrow disks. Although it is possible that giant impacts occurring at $t > 20$ Myr might occur as outliers in truncated, depleted, or peaked disks, we believe that more detailed investigations are needed to reveal the disk conditions that can allow late giant impacts to occur systematically.

*E. Earth's late accreted mass* – As our results revealed that the late veneer mass is inversely correlated with the timing of the last giant impact experienced by Earth (constraint D) (this result is also in agreement with that found by Jacobson & Morbidelli 2014), we obtained quite low success rates for this constraint. In fact, our Earth analogs typically acquired late veneer mass fractions larger than the upper limit of ~2% of the Earth's mass by a factor of 2–10. New investigations concerning Earth's last giant impact will probably satisfy both constraints D and E.

*F. Formation time of Mercury, Venus, Earth, and Mars* – In general, truncated, depleted, and peaked disks formed the bulk of their planet analogs in timescales compatible with that inferred for the terrestrial planets. In particular, even the rapid formation of Mars was not an issue in the majority of the obtained systems. However, the results were less successful in peaked disks that considered disk mass distributions more extended than those used in truncated/depleted disks.

*G. Origin of water on Mercury, Venus, Earth, and Mars* – Overall, considering the uncertainties in the WMFs of Mercury, Venus, and Mars, all disk models were able to produce analogs of these planets with acceptable water masses. However, because the Earth analogs acquired their bulk masses from the initially dry core regions in truncated and depleted disks, these planets were too dry in general.

Moreover, the Earth analogs obtained in fiducial and peaked disks could barely reach Earth's estimated minimum WMF ($2.5 \cdot 10^{-4}$). Assuming that all objects beyond 2.5 au were water-rich with an initial WMF = 10% (e.g., Clement et al. 2018, 2019a) does not solve the problem, because the Earth analogs accreted only ~1% of their final masses from that region. This is not surprising, as such a water model would be very similar to that adopted in this work. However, the inclusion of outer regions in depleted and peaked disks increased the chances of Earth analogs to acquire sufficient WMFs (Table 3). Alternatively, if planetesimals in the outer region of truncated disks were water-rich (beyond ~1–1.2 au), as assumed in O'Brien et al. (2014), the Earth analogs could acquire enough water (Table 6). However, the same assumption would imply excessively wet Venus and Mars analogs (Tables 5 and 7). These results suggest that water delivery to Earth must be revisited and that the same delivery should be considered for all terrestrial planets simultaneously.

*H. Orbital structure of the asteroid belt* – Our simulations were not designed to investigate the asteroid belt, and thus we only discuss the results for our seventeen 4-P systems (Section 3.4). First, the asteroid belts obtained in truncated disks (systems #1–11) essentially reflect their initial conditions, such that they lack the low-$e$ component of the real asteroid belt. Nevertheless, the dynamical evolution of the giant planets could reproduce that component at a later stage (e.g., Deienno et al. 2018). The four asteroid belts formed in systems #12–14 (depleted disks) and #17 (peaked disk) acquired a similarly wide range of eccentricities and inclinations at the end of 400 Myr, resembling the orbital distribution in the asteroid belt. The asteroid belts in systems #15 and #16 were strongly perturbed by the icy massive embryos in the I-belts, such that their final $e$–$i$ distribution is quite different from that in the real belt. For these reasons, depleted-IB disks lack the potential to reproduce the orbital structure of the asteroid belt.

### 4.1 Origin of Mercury

The properties of our representative Mercury analogs indicate that disks with core regions at 0.7–1.0(1.2) au, as modeled in truncated and depleted disks, are unable to reproduce Mercury's dynamically excited orbit at ~0.4 au *and* its low mass at the same time (see Sections 3.2.3, 3.3.1, and 3.4 for details). Furthermore, the Mercury analogs in truncated disks did not experience giant impacts. If such giant impacts are required to explain Mercury's high bulk density (Asphaug & Reufer 2014; Jackson et al. 2018), then this disk model is further disfavored. Perhaps truncated disks as modeled in Hansen (2009) with large numbers of smaller embryos could increase the odds of giant impacts, but the probability of success appears to be quite low, e.g., less than 1/38 (<3%)[8]. Moreover, N-body simulations of Mercury's collisional origin that considered such truncated disks and other disks akin to our D7-12 disks could not reproduce Mercury (Clement et al. 2019c). For these reasons, the origin of Mercury as a core region embryo that was scattered to its current orbit seems extremely unlikely in narrow disks in general. Even setting core regions with smaller inner edges will not solve the problem. Such disks may produce Mercury analogs with orbits closer to the observed $a$ ~ 0.4 au, but they will likely produce Venus analogs located too close to the Sun (i.e., near the location of the edge, as found in Section 3.2.3).

---

[8] SPH models for the collisional origin of Mercury suggest that proto-Mercury was up to ~4 times more massive than Mercury (Asphaug & Reufer 2014; Chau et al. 2018). The single representative Mercury analog in Hansen (2009) (their Sim1 out of 38 simulations) experienced giant impacts, but its final mass was smaller than that of Mercury.

The Mercury analogs obtained in disks with inner regions probably have better chances of reproducing Mercury. Despite the low statistics, two Mercury analogs obtained in peaked disks satisfied both the planet's orbit and mass constraints. These analogs also acquired their bulk masses via three to four giant collisions with other similar-mass embryos, a collisional history that might be compatible with the multiple hit-and-run impacts scenario of Asphaug & Reufer (2014) (see also Jackson et al. 2018 and Clement et al. 2019c).

Mercury analogs also acquired late veneer mass via impacts of planetesimals from the outer regions in depleted and peaked disks. In agreement with the findings of LI17, disks with such components could explain the presence of water/volatiles in Mercury.

In conclusion, given the difficulties discussed above, detailed exploration of disks containing both inner and outer region components or perhaps more sophisticated disk models are warranted.

### 4.2 Origin of Mars

First, as detailed in Sections 3.3.4 and 3.4, the properties of our representative Mars analogs suggest that only truncated disks are unable to reproduce the moderately excited orbit *and* low mass of Mars simultaneously. In particular, because of the dynamical friction imparted by planetesimals and the lack of excitation mechanisms in the Mars region (e.g., no embryos, giant planets on low-$e$ orbits), the Mars analogs acquired medians $e = 0.01$–$0.02$, $i = 1$–$2$ deg for truncated disk subsets T-r1/r4/r8. Although these analogs were scattered into the Mars region within the first Myr of evolution, their orbits were damped in only 10–20 Myr timescales via dynamical friction. These analogs could acquire more excited orbits during the subsequent giant planet migration/instabilities, but past studies have shown that reproducing Mars's orbit is unlikely in that scenario (Brasser et al. 2013; Kaib & Chambers 2016; Clement et al. 2018, 2019a). Nevertheless, Mars's current $e/i$ might be the result of chaotic dynamics (Laskar 2008). Finally, fiducial disks were highly inefficient at producing Mars analogs and the analogs produced were too dynamically excited.

In contrast, depleted disks modeled with $r = 4$ yielded Mars analogs with medians $a = 1.6$ (1.5) au, $e = 0.08$ (0.05), $i = 3.7$ (2.9) deg, and $m = 0.18$ (0.21) $M_\oplus$ for D-r4 (D-r1) disks. Peaked disks produced seven low-mass representative Mars analogs with medians $a = 1.6$ au and $m = 0.07$ $M_\oplus$, but their orbits were somewhat too cold ($e = 0.03$, $i = 2.4$ deg) and only two of the analogs formed within 10 Myr. However, new setups of peaked disks with reduced dynamical friction in the Mars region (e.g., using larger $r$) and different mass distributions beyond ~1 au may remedy these problems. Overall, our results suggest that reproducing Mars would require the following conditions: 1) dynamically active disks containing outer regions populated with small embryos and/or the Jupiter–Saturn pair initially on eccentric orbits, and; 2) a disk in which most of the mass was distributed in embryos rather than planetesimals.

Finally, out of 173 analog systems, 114, 53, and 6 systems formed a single, two, and three Mars analogs, respectively. These results support the idea that the inner solar system once contained a "Planet V" (Chambers 2007) or even a "Planet VI" at the end of the terrestrial planet formation. Perhaps such additional Mars analogs collided with another planet/the Sun or were dynamically ejected from the solar system at some point in the solar system history. Indeed, a "Goldilocks" giant planet migration/instability would be required to selectively remove such additional analogs.

### 4.3 Existence of I-belts and planet analog mass ratios

The perturbation of massive icy embryos placed at 3.4–4.0(4.3) au within I-belts reduced the chances of forming Mars analogs and produced overly dynamically excited systems of planets and asteroids. Another drawback was that such massive embryos could survive as unwanted planets in the asteroid belt after 400 Myr of evolution. Finally, the increase in the WMFs acquired by the planet analogs was only modest when compared to depleted-only disks. For these reasons, I-belts probably did not exist at the time of terrestrial planet formation.

As detailed in Table 3, reproducing the mass ratios of Mercury–Venus, Venus–Earth, and Mars–Earth is still difficult to achieve consistently. Even the 0.82 mass fraction of Venus–Earth remains a challenge for all disks. Nevertheless, disks with outer region components (depleted and peaked) appear to produce better Venus–Earth mass fractions, because the feeding zones of the Earth analogs extend beyond the core region. Overall, peaked disks yielded the best mass ratios compared with the other disk models tested, but more detailed modeling is needed to draw a firmer conclusion.

## 5. SUMMARY

We performed extensive simulations of terrestrial planet formation that allowed us to constrain the protoplanetary disk conditions that favored the formation of systems analogous to our own. In particular, we investigated the influence of disk core size, inner and outer region disk components, different distributions of disk mass in embryos/planetesimals, and distinct giant planet orbital architectures. We also developed a classification algorithm to accurately identify planet analogs in a given system, which then allowed us to determine planetary systems that seemed to resemble the inner solar system (terrestrial planet analog systems).

We found a total of 194 terrestrial planet analog systems, 17 of which contained representative analogs of the four terrestrial planets within the same system. However, while our results showed that forming Venus, Earth, and Mars is relatively easy (173 out of the 194 analog systems), obtaining planets analogous to Mercury is rare: only 38 were found among the 194 analog systems. Additionally, the obtained analogs of Mercury and Mars are commonly more massive, too dynamically cold, and formed too close to Venus and Earth in comparison with the real planets. Finally, the analogs of Earth were in general too dry, experienced excessively early Moon-forming giant impacts, and accreted too much mass after those impacts. As these problems were systematic in truncated disks just as overly massive Mars-like planets were in fiducial disks, we conclude that neither disk model can explain the formation of the terrestrial planets of the solar system. This means that truncated disks based on typical outcomes of the Grand Tack model and the often assumed 0.7–1.0 au disk in the Empty Asteroid Belt model are strongly disfavored.

Despite these difficulties, based on the best prospects to satisfy the fundamental constraints of the inner solar system, a detailed analysis of the results allowed us to identify the most favorable properties for the protoplanetary disk (key justifications follow each item):

- Mass concentrated in narrow core regions between ~0.7–0.9 and ~1.0–1.2 au: More analog systems in general (thanks to higher production of Mercury and Mars analogs).

- An inner region component starting at ~0.3–0.4 au: More Mercury analogs (including those with low-mass and excited orbits).

- An outer region component starting at ~1.0–1.2 au and much less massive than the core region: Systematic delivery of water to Mercury and higher chances of satisfying Earth's water budget, better Earth–Venus mass ratios, production of Ceres-like objects, asteroid belts with orbital structure similar to observations.

- Embryos rather than planetesimals carrying most of the disk mass: Better analog systems for disk models that considered $r = 4$ or $8$ (rather than $r = 1$), more Mars analogs with moderately excited orbits, more four-terrestrial-planet analog systems.

- Jupiter and Saturn placed on eccentric orbits (even if locked in a mutual MMR): Mars analogs that are less massive, fewer planets stranded in the asteroid belt.

Some of these properties are present in the depleted and peaked disks, which might represent the typical outcomes of the Early Instability and Pebble Accretion models. However, it is unclear whether any of these models can simultaneously produce the initial disk conditions summarized above. Furthermore, reproducing the locations of Mercury and Venus will require some model fine-tuning, such as exploring disks with different disk inner edge locations and core sizes. Reproducing Earth's WMF of a few ocean masses, last giant impact later than ~20 Myr (Moon formation), late veneer mass fraction less than ~0.02 $M_\oplus$, as well the formation of a single Mars at its current location will require further exploration of disk models with core regions surrounded by mass-depleted inner and outer region components. Finally, reproducing the small masses and relatively excited orbits of both Mercury and Mars will likely require more dynamically active protoplanetary disks (with large $r$, giant planets on eccentric orbits, embryos placed across the entire disk, etc.).

Our results revealed a number of protoplanetary disk conditions that may be required to reproduce the four terrestrial planets consistently. Therefore, new models of planetesimal or giant planet formation that can produce such disk conditions are warranted. Future models of terrestrial planet formation will also reveal the optimal disk conditions needed to satisfy all constraints in the inner solar system, thus offering new insights into the origin of Ceres, the formation of the asteroid belt, and planetesimal/embryo accretion.

## ACKNOWLEDGMENTS

We would like to thank very much the referee for a number of helpful and detailed comments, which allowed us to improve the overall presentation and flow of this work. We also appreciate comments by H. Levison (about densities of forming planets) and discussion with D. P. O'Brien and K. Kretke about disk initial conditions during the preparation of this work. All simulations presented in this work were performed using the general-purpose PC cluster at the Center for Computational Astrophysics (CfCA) in the National Astronomical Observatory of Japan (NAOJ). We are thankful for the generous time allocated to run the simulations. T.I. acknowledges research funding from the JSPS Kakenhi Grant, JP16K05546/2016-2018 and JP18K03730/2018-2021.

# APPENDIX

We provide further details about the planet analogs obtained in our 17 Mercury-Venus-Earth-Mars analog systems (4-P analog systems), as discussed in Section 3.4. Figure A1 illustrates the feeding zones of all planet analogs and Table A1 summarizes a number of key variables for these planets.

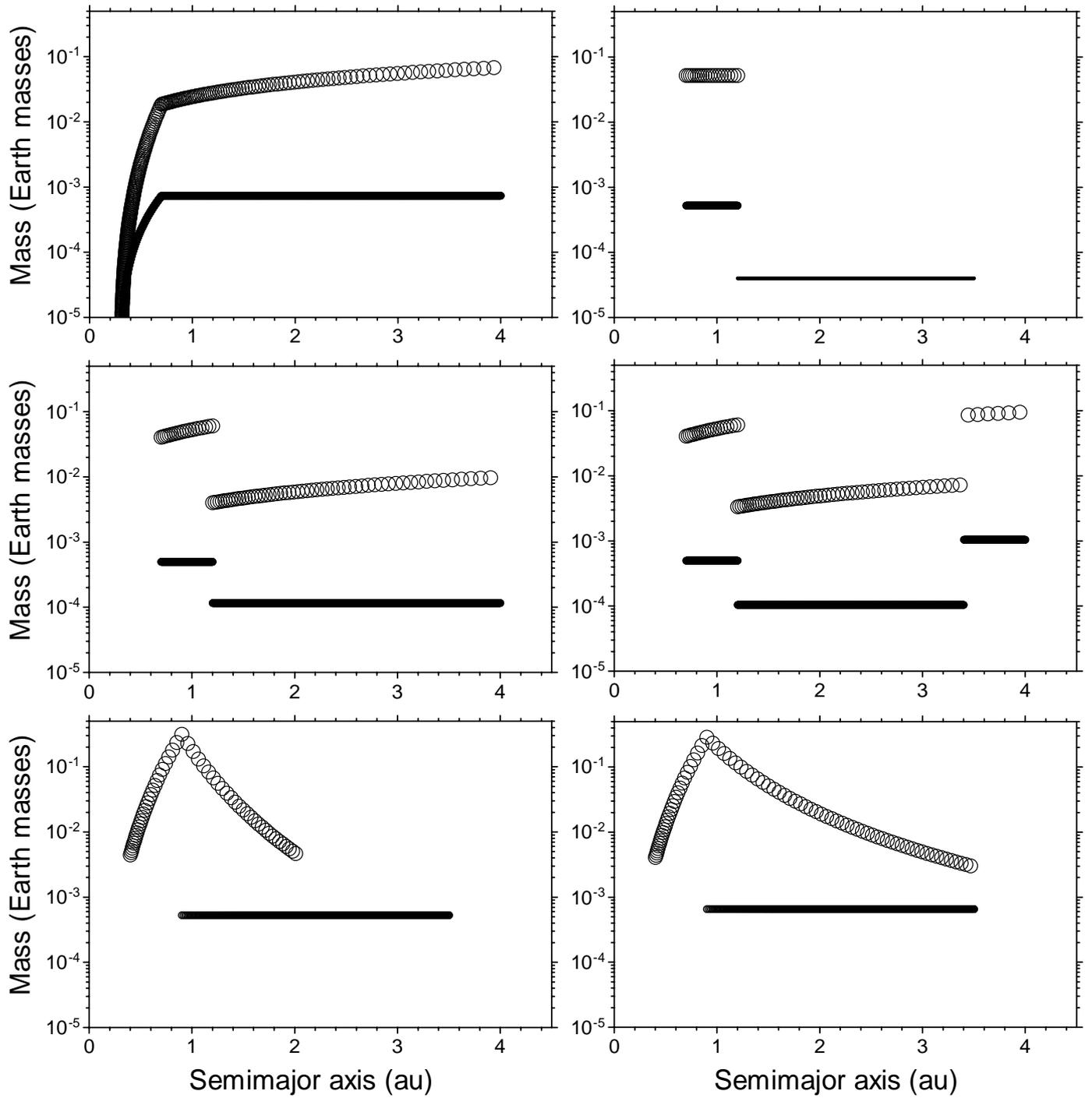

**Figure 1**. Representative initial conditions of embryos (large symbols) and planetesimals (small symbols) for the four protoplanetary disks modeled in this work: fiducial, truncated [top]; depleted-only, depleted-IB [middle]; peaked (P9-20ir4), peaked (P9-35ir4) [bottom]. See Table 2 and main text for details.

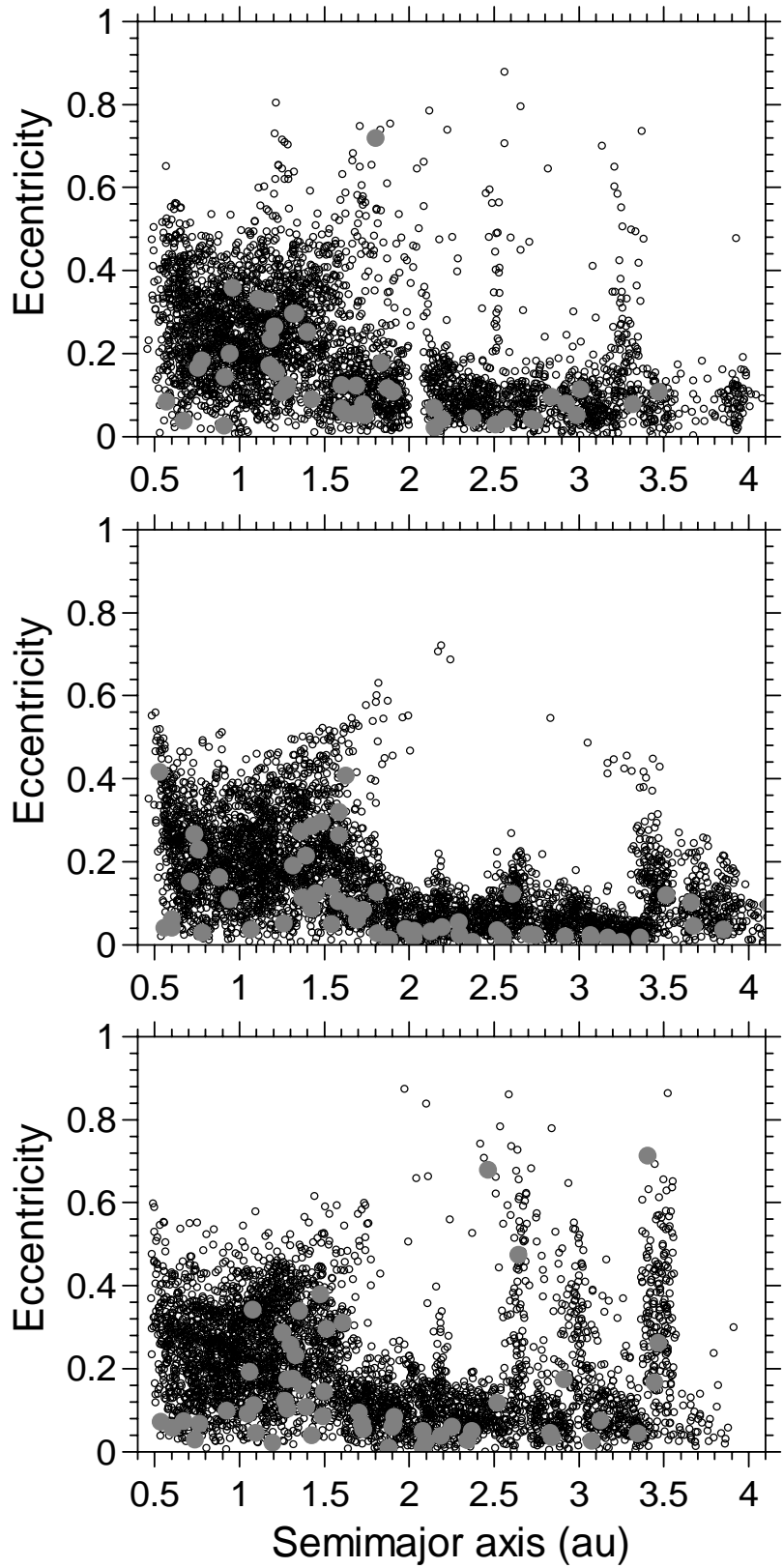

**Figure 2**. Orbital state of systems representative of the NC [top], JS23me [middle], and JS12he [bottom] orbital architectures of the giant planets after 3 Myr of evolution. Embryos and planetesimals are represented by gray and black symbols, respectively. The three examples came from D7-10 depleted disks. See Tables 1, 2, and Section 3.2.2 for details.

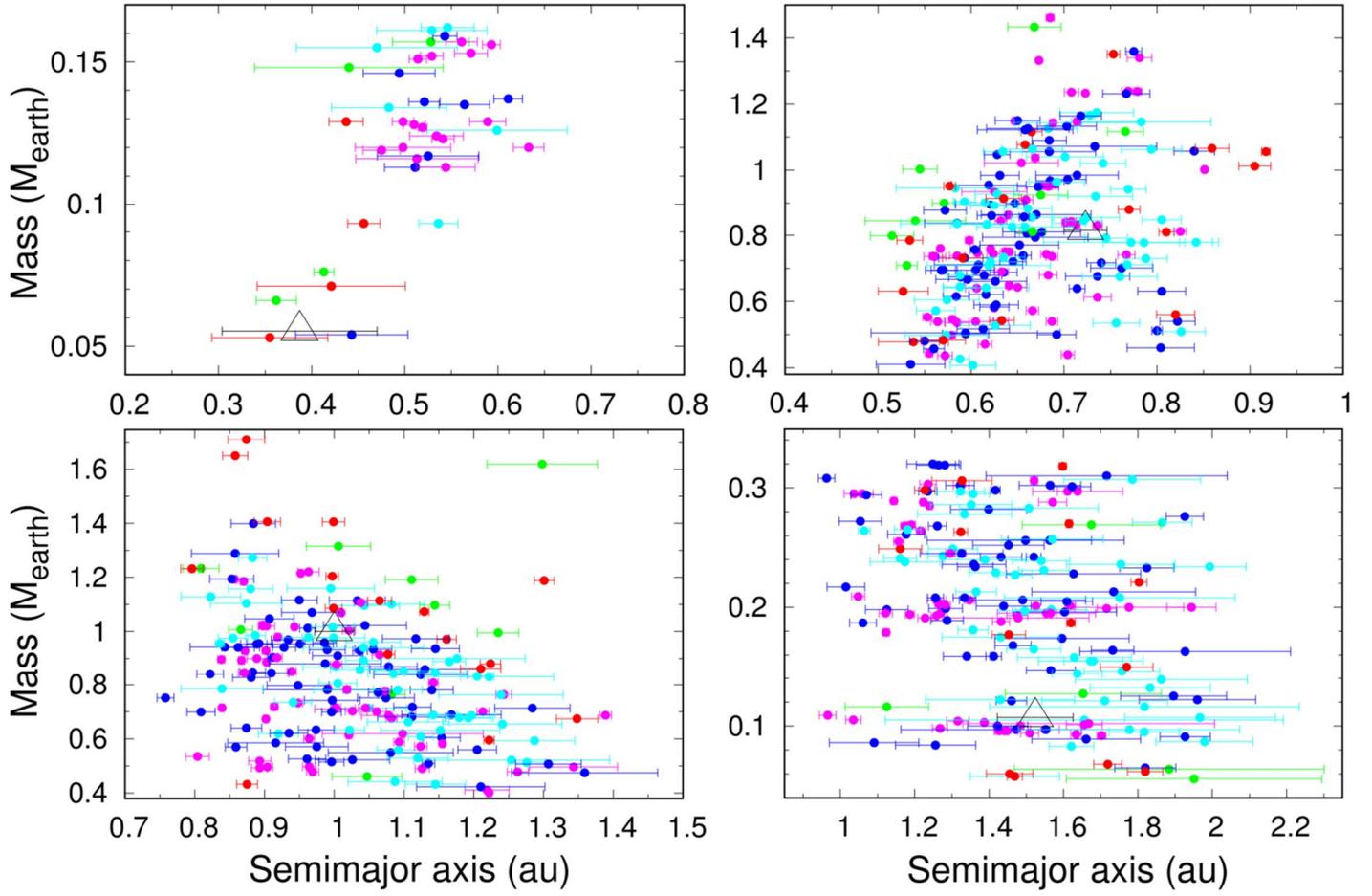

**Figure 3**. Representative analogs obtained in three-/four-terrestrial-planet analog systems: 38 Mercury analogs [top left], 194 Venus analogs [top right], 194 Earth analogs [bottom left], and 173 Mars analogs [bottom right]. Further details are described in Tables 4–7. The different colors indicate the disk model used: fiducial (green), truncated (magenta), depleted-only (blue), depleted-IB (cyan), and peaked (red). The large open triangles represent the solar system terrestrial planets.

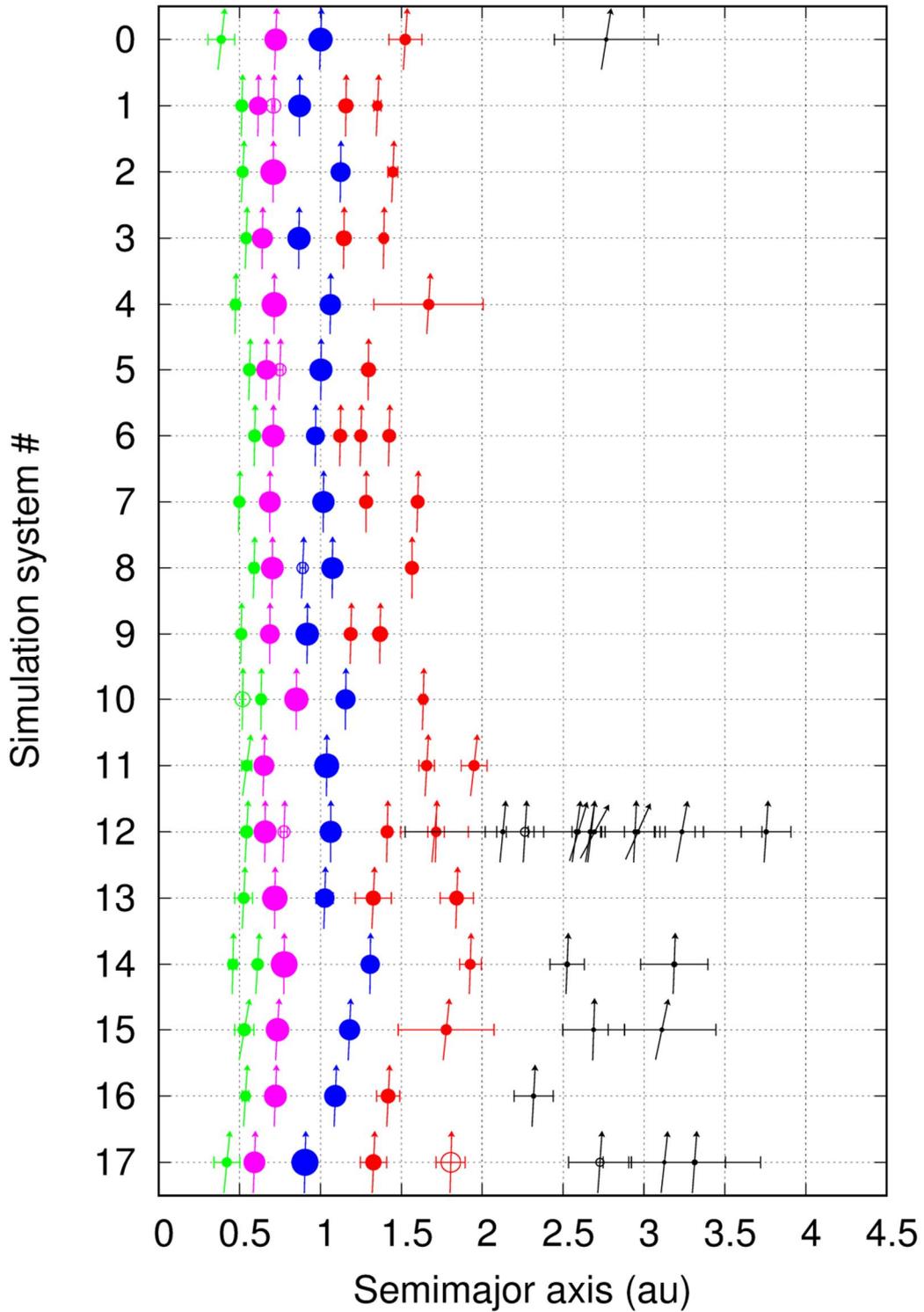

**Figure 4.** Comparison of individual 4-P analog systems (#1–17) obtained after 400 Myr of dynamical evolution with the solar system planets and Ceres (shown as "system #0" at the top). The data are combined according to fiducial (none), truncated (#1–11), depleted-only (#12–14), depleted-IB (#15–16), and peaked (#17) protoplanetary disks. The inclination of the system objects is represented by the angle between the vector and the perpendicular (e.g., the vector would point to the top (right) if the object's $i = 0$ (90) deg). Error bars represent the range of heliocentric distance based on the object's perihelion and aphelion. The size of a planet scales proportional to its mass to the power (1/3). See Section 3.4 for details.

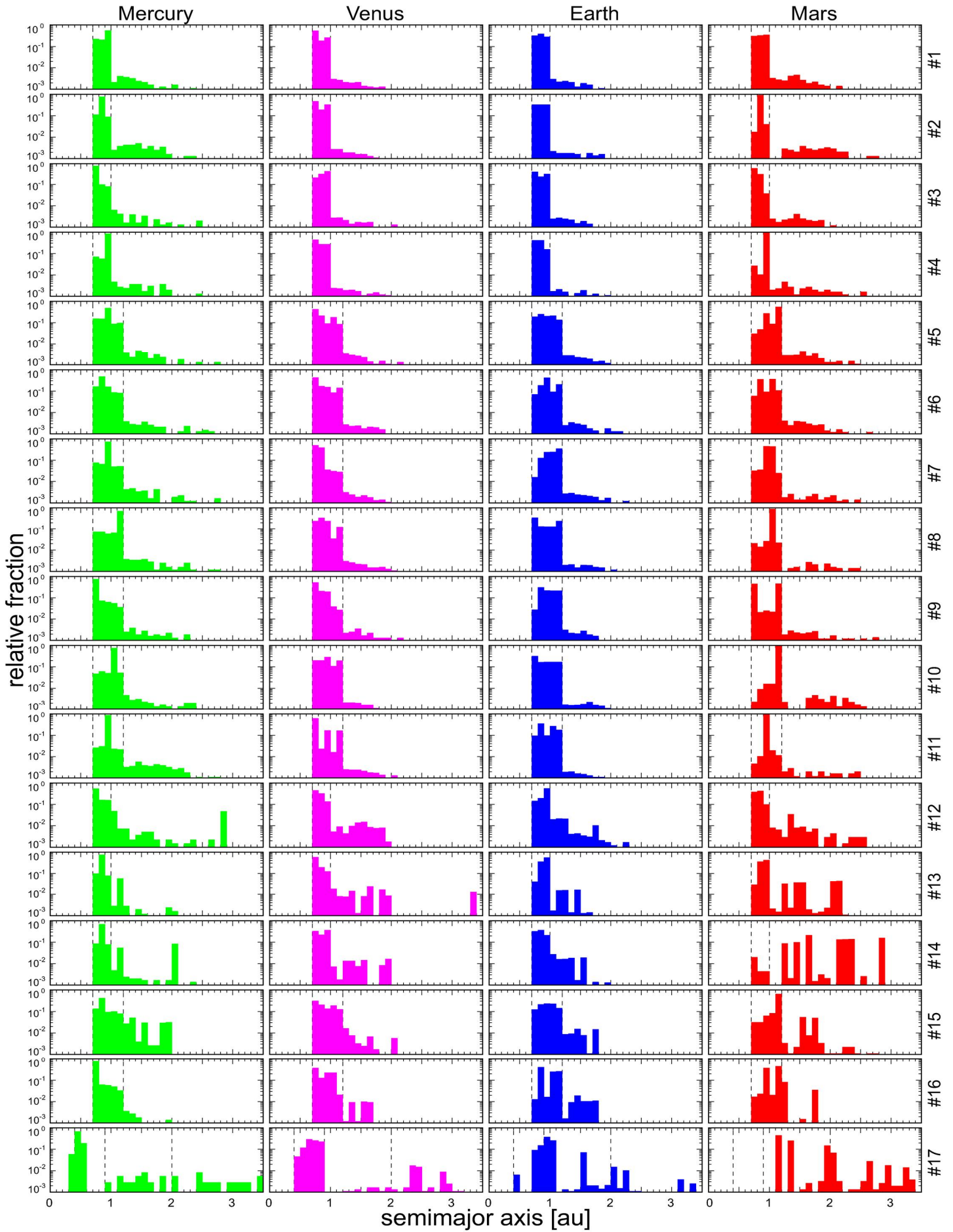

**Figure A1**. Feeding zones of Mercury, Venus, Earth, and Mars representative analogs obtained in 17 4-P analog systems (Figure 4), ordered from top to bottom panels.

**Table 1**. Initial conditions for the giant planets in our simulations

| Orbital Architecture | $a_{J0}$ (au) | $e_{J0}$ | $i_{J0}$ (deg) | $a_{S0}$ (au) | $e_{S0}$ | $i_{S0}$ (deg) |
|---|---|---|---|---|---|---|
| 2:3 MMR (JS23) [a] | 5.513 | 0.004 | 0.1 | 7.456 | 0.015 | 0.1 |
| 2:3 MMR moderate-*e* (JS23me) | 5.511 | 0.020 | 0.5 | 7.261 | 0.071 | 1.4 |
| 1:2 MMR high-*e* (JS12he) | 5.507 | 0.083 | 0.1 | 8.763 | 0.154 | 0.2 |
| Near-current (NC) | 5.230-5.267 | 0.065-0.11 | 1.5 | 9.585-9.595 | 0.065-0.11 | 2.5 |

**Notes.** Subscripts J and S refer to Jupiter and Saturn, respectively. JS23, JS23me, and JS12he orbital architectures started with Jupiter and Saturn locked in a mutual *p:q* mean motion resonance (MMR), whereas the NC orbital architecture started with Jupiter and Saturn on orbits close to their current orbits. In this particular architecture, slightly higher eccentricities were used in order for Jupiter and Saturn to acquire nearly their current eccentricities after suffering dynamical friction from embryos/planetesimals over the simulated 400 Myr. See Section 2 for more details.

[a] In systems with the four giant planets (JS23-4GPs), Uranus and Neptune started with $a$ = 9.734 au, $e$ = 0.042, $i$ = 0.2 deg and $a$ = 11.867 au, $e$ = 0.010, $i$ = 0.1 deg, respectively. The four giant planets were locked in a 2:3-2:3-4:3 MMR chain.

**Table 2.** Initial conditions for the protoplanetary disks in our simulations

| Disk model | Disk | Inner Region (au) [mass ($M_\oplus$)] | Core Region (au) [mass ($M_\oplus$)] | Outer Region (au) [mass ($M_\oplus$)] | Number of Embryos | Number of Planetesimals | Giant Planet Orbital Architecture | Ratio $r$ |
|---|---|---|---|---|---|---|---|---|
| fiducial [a] | F5 | - | 0.5-4.0 [6.0] | - | 116 | 4659 | JS12he, NC | 1 |
| | | - | 0.5-4.3 [6.3] | - | 119 | 4925 | JS23, JS23me | 1 |
| | F7 | - | 0.7-4.0 [8.2] | - | 77 | 4192 | JS12he, NC | 1 |
| | | - | 0.7-4.3 [8.6] | - | 79 | 4458 | JS23, JS23me | 1 |
| | F7ir3 | 0.3-0.7 [0.5] | 0.7-4.0 [6.2] | - | 150 | 5384 | JS12he, NC | 1 |
| | | 0.3-0.7 [0.5] | 0.7-4.3 [6.6] | - | 153 | 5650 | JS23, JS23me | 1 |
| truncated [b] | T7-10 | - | 0.7-1.0 [2.1] | 1.0-3.5 [0.2] | 20 | 7000 | JS23, JS23-4GPs | 1,4,8 |
| | T7-12 | - | 0.7-1.2 [2.1] | 1.2-3.5 [0.2] | 20 | 7000 | JS23, JS23-4GPs | 1,4,8 |
| depleted [c] | D7-10 | - | 0.7-1.0 [2.0] | 1.0-4.0 [0.7] | 74-80 | 5000 | JS12he, NC | 1,4 |
| | | - | 0.7-1.0 [2.0] | 1.0-4.3 [0.7] | 77-82 | 5000 | JS23me | 1,4 |
| | D7-12 | - | 0.7-1.2 [2.0] | 1.2-4.0 [0.7] | 74-80 | 5000 | JS12he, NC | 1,4 |
| | | - | 0.7-1.2 [2.0] | 1.2-4.3 [0.7] | 76-81 | 5000 | JS23me | 1,4 |
| | D7-10IB | - | 0.7-1.0 [2.0] | 1.0-4.0 [0.3+0.5] | 73-80 | 5000 | JS12he, NC | 1,4 |
| | | - | 0.7-1.0 [2.0] | 1.0-4.3 [0.3+0.8] | 76-81 | 5000 | JS23me | 1,4 |
| | D7-12IB | - | 0.7-1.2 [2.0] | 1.2-4.0 [0.3+0.5] | 72-77 | 5000 | JS12he, NC | 1,4 |
| | | - | 0.7-1.2 [2.0] | 1.2-4.3 [0.3+0.8] | 76-80 | 5000 | JS23me | 1,4 |
| peaked [d] | P9-20ir4 | 0.4-0.9 [1.3] | 0.9-2.0 [1.8,1.6] | 2.0-3.5 [0.9,0.3] | 59 | 2230 | JS23-4GPs | ~6,19 |
| | P9-20ir4#2 | 0.4-0.9 [1.1] | 0.9-2.0 [1.7,1.5] | 2.0-3.5 [0.9,0.3] | 66 | 2230 | JS23-4GPs | ~6,19 |
| | P9-35ir4 | 0.4-0.9 [1.2] | 0.9-2.0 [2.2,2.0] | 2.0-3.5 [1.4,0.6] | 82 | 2230 | JS23-4GPs | ~6,19 |

**Notes.** $r$ represents the ratio of total disk mass in embryos to that in planetesimals. The total number of simulation runs per disk = number of giant planet orbital architectures times the number of ratios $r$ tested. The number of embryos varied slightly depending on the $r$ value used. The giant planet orbital architectures are shown in Table 1. See Section 2 for more details about the disk models.

[a] The disk outer edge was determined by the condition $a_J(1 - e_J) - 3r_{H,J}$, where $r_{H,J}$ is the Hill radius, $a_J$ is the semimajor axis, and $e_J$ is the eccentricity of Jupiter at the start of the simulations.

[b] We placed 20 embryos and 2000 planetesimals ("S-type planetesimals") in the core region and 5000 planetesimals in the outer region ("C-type planetesimals"). A disk outer edge of 3.5 au was assumed for the C-planetesimals.

[c] We placed 20 embryos and 2000 planetesimals in the core region and the remaining embryos and planetesimals in the outer region, which was mass-depleted compared with the former region. Disks D7-10(12)IB possess a massive component (nondepleted) located in between 3.4 au and 4.0 (or 4.3) au. For these disks in particular, the masses in the depleted and IB parts of the disk in the outer region are represented by a sum of two numbers, respectively. The disk outer edge was defined in the same way as for fiducial disks.

[d] Disks P9-20ir4, P9-20ir4#2, and P9-35ir4 have distinct disk mass distributions. Disk P9-20ir4#2 has the embryos distributed with steeper mass distributions around the 0.9 au point compared to disk P9-20ir4. We placed the embryos across the inner and core regions in disks P9-20ir4 and P9-20ir4#2 (the entire disk in disk P9-35ir4), and the planetesimals across the core and outer regions in all disks. The total mass in the core and outer regions varies depending on the value of $r$ considered. As for truncated disks, a disk outer edge of 3.5 au was assumed here.

**Table 3.** Summary of key variables obtained for the three-/four-terrestrial-planet analog systems

| Disk Model | N0 | n3$_{mve}$ | n3$_{vem}$ | n4 | AMD | RMC | OS | Mpf | $m_{Ma}$ (M$_\oplus$) | Lvf (%) | MVr | VEr | MEr | C1 (%) | C2 (%) | C3 (%) | C4 (%) | C5 (%) | C6 (%) | C7 (%) | C8 (%) |
|---|---|---|---|---|---|---|---|---|---|---|---|---|---|---|---|---|---|---|---|---|---|
| fiducial | 120 | 4 | 5 | 0 | 0.0025 | 34.9 | 32.5 | 1.0 | 0.12 | 7 | 0.12 | 0.82 | 0.11 | 67 | 33 | 100 | 78 | 67 | 22 | 44 | 22 |
| truncated | 120 | 5 | 42 | 11 | 0.0002 | 65.0 | 24.1 | 1.5 | 0.20 | 9 | 0.17 | 1.04 | 0.27 | 98 | 88 | 59 | 98 | 21 | 10 | 0 | 95 |
| *T7-10 | 60 | *2* | *20* | *4* | *0.0002* | *70.1* | *27.1* | *1.3* | *0.20* | *9* | *0.17* | *1.00* | *0.26* | *100* | *88* | *73* | *100* | *15* | *4* | *0* | *96* |
| *T7-12 | 60 | *3* | *22* | *7* | *0.0002* | *62.6* | *18.1* | *1.7* | *0.20* | *6* | *0.17* | *1.03* | *0.28* | *97* | *88* | *47* | *97* | *25* | *16* | *0* | *94* |
| *T-r1 | 40 | *2* | *8* | *3* | *0.0001* | *78.1* | *16.4* | *2.0* | *0.24* | *26* | *0.22* | *0.96* | *0.34* | *100* | *100* | *38* | *100* | *8* | *0* | *0* | *92* |
| *T-r4 | 40 | *1* | *18* | *6* | *0.0002* | *63.6* | *17.9* | *1.5* | *0.19* | *9* | *0.16* | *1.14* | *0.27* | *100* | *88* | *48* | *96* | *16* | *4* | *0* | *100* |
| *T-r8 | 40 | *2* | *16* | *2* | *0.0004* | *54.0* | *32.4* | *1.3* | *0.20* | *4* | *0.14* | *0.99* | *0.23* | *95* | *80* | *85* | *100* | *35* | *25* | *0* | *90* |
| depleted-only | 120 | 5 | 53 | 3 | 0.0016 | 53.5 | 29.8 | 1.4 | 0.21 | 12 | 0.18 | 0.88 | 0.25 | 79 | 92 | 80 | 97 | 30 | 16 | 26 | 92 |
| *D7-10 | 60 | *4* | *28* | *3* | *0.0013* | *55.8* | *29.8* | *1.4* | *0.21* | *13* | *0.17* | *0.96* | *0.25* | *77* | *94* | *86* | *94* | *29* | *17* | *20* | *89* |
| *D7-12 | 60 | *1* | *25* | *0* | *0.0019* | *51.6* | *29.9* | *1.4* | *0.24* | *10* | *0.19* | *0.89* | *0.30* | *81* | *88* | *73* | *100* | *31* | *15* | *35* | *96* |
| *D-r1 | 60 | *3* | *26* | *1* | *0.0013* | *55.2* | *26.1* | *1.6* | *0.20* | *22* | *0.19* | *0.95* | *0.27* | *80* | *90* | *70* | *97* | *20* | *17* | *23* | *90* |
| *D-r4 | 60 | *2* | *27* | *2* | *0.0018* | *52.7* | *31.8* | *1.2* | *0.23* | *9* | *0.16* | *0.87* | *0.25* | *77* | *94* | *90* | *97* | *39* | *16* | *29* | *94* |
| *D-NC | 40 | *2* | *19* | *0* | *0.0024* | *70.4* | *30.5* | *1.2* | *0.17* | *10* | *0.14* | *0.85* | *0.21* | *67* | *100* | *100* | *100* | *38* | *19* | *0* | *86* |
| *D-JS23me | 40 | *2* | *16* | *2* | *0.0013* | *49.2* | *19.6* | *1.6* | *0.26* | *15* | *0.17* | *0.94* | *0.34* | *80* | *85* | *55* | *95* | *15* | *15* | *30* | *100* |
| *D-JS12he | 40 | *1* | *18* | *1* | *0.0012* | *52.8* | *30.2* | *1.3* | *0.20* | *9* | *0.17* | *0.97* | *0.23* | *90* | *90* | *85* | *95* | *35* | *15* | *50* | *90* |
| depleted-IB | 120 | 4 | 43 | 2 | 0.0018 | 55.5 | 30.9 | 1.2 | 0.21 | 5 | 0.17 | 1.01 | 0.25 | 71 | 94 | 84 | 69 | 33 | 35 | 41 | 92 |
| peaked | 60 | 3 | 13 | 1 | 0.0007 | 38.8 | 20.1 | 1.4 | 0.20 | 12 | 0.09 | 0.81 | 0.19 | 100 | 18 | 53 | 71 | 18 | 12 | 41 | 59 |
| 4-P systems | 540 | - | - | 17 | 0.0002 | 68.4 | 17.8 | 1.5 | 0.20 | 10 | 0.15 | 1.09 | 0.25 | 94 | 94 | 29 | 100 | 18 | 6 | 6 | 94 |
| **Solar system** | - | - | - | - | **0.0018** | **89.7** | **37.7** | - | **0.11** | **0.1-2** | **0.07** | **0.82** | **0.11** | - | - | - | - | - | - | - | - |

**Notes.** N0 is the number of simulation runs performed for a given disk model or a subset of it (in italics). Disk model subsets D/T-r<number> refer to systems combined by a common ratio $r$, where <number> represents the value of $r$ used. Subsets D-<GP> refer to systems combined by a common giant planet orbital architecture, where <GP> represents the architecture considered (Table 1). 4-P systems represent analog systems in which the four terrestrial planet representative analogs are present in the system. 4-P systems are taken altogether when determining the various quantities, irrespective of the disk model from which these systems originated. The number of analog systems containing Mercury–Venus–Earth, Venus–Earth–Mars, and Mercury–Venus–Earth–Mars representative analogs are given by n3$_{mve}$, n3$_{vem}$, and n4, respectively. Conversely, the total number of analog systems is given by NAS = n3$_{mve}$ + n3$_{vem}$ + n4. AMD is the angular momentum deficit, RMC is the radial mass concentration, OS is the orbital spacing, Mpf is the Mars analog production factor (defined by the total number of Mars analogs produced divided by number of systems containing at least one Mars analog), $m_{Ma}$ is the median mass of the representative Mars analogs, Lvf is the late veneer mass fraction, MVr is the ratio of the Mercury analog mass to the Venus analog mass, VEr is the ratio of the Venus analog mass to the Earth analog mass, and MEr is the ratio of the Mars analog mass to the Earth analog mass. In general, the earlier the timing of the last giant impact suffered by the Earth analog (tGI), the larger the mass accreted after that impact (Lvf). Here, we found that Lvf > 10% reflects the median tGI < 20 Myr. The last eight columns represent the fractions of NAS that satisfied a particular constraint, as defined by C1: AMD (0–0.0036), C2: RMC (44.9–179.4), C3: OS (18.9–75.4); C4: unwanted planets in the asteroid belt (maximum one object less massive than Mars); C5: tGI for the Earth analog (20–140 Myr); C6: Lvf for the Earth analog (0.1–2%); C7: WMF for the Earth analog (2.5·10$^{-4}$–1·10$^{-2}$); C8: Mars analog formation time (within 10 Myr). See Section 1.1 for more details on these quantities.

**Table 4.** Representative Mercury analogs obtained in three-/four-terrestrial-planet analog systems

| | N | $a$ (au) | $e$ | $i$ (deg) | $m$ ($M_\oplus$) | WMF | nGI | tGI (Myr) | $f_0$ | $f_{Emb}$ | $f_{Obj}$ | $a_{0\_mass}$ (au) | | $a_{0\_H2O}$ (au) | |
|---|---|---|---|---|---|---|---|---|---|---|---|---|---|---|---|
| all | 38 | 0.52 | 0.05 | 2.7 | 0.13 | $2\cdot10^{-4}$ | 0 | 21 | 0.45 | 0.11 | 0.44 | 0.7 | 1.2 | 2.0 | 3.0 |
| fiducial | 4 | 0.43 | 0.07 | 3.8 | 0.11 | $3\cdot10^{-2}$ | 4 | 19 | 0.08 | 0.10 | 0.82 | 0.3 | 4.0 | 3.8 | 4.0 |
| truncated | 16 | 0.53 | 0.03 | 1.3 | 0.13 | $2\cdot10^{-4}$ [b] | 0 | - | 0.60 | 0.00 | 0.40 | 0.7 | 1.1 | 2.0 | 3.1 |
| depleted | 8 | 0.52 | 0.06 | 2.9 | 0.14 | $2\cdot10^{-4}$ | 0 | 67 | 0.46 | 0.10 | 0.44 | 0.7 | 1.2 | 2.0 | 2.9 |
| depleted-IB | 6 | 0.53 | 0.12 | 5.4 | 0.14 | $2\cdot10^{-5}$ | 1 | 37 | 0.43 | 0.13 | 0.43 | 0.7 | 1.4 | 1.5 | 2.5 |
| peaked | 4 | 0.43 | 0.11 | 4.8 | 0.08 | $1\cdot10^{-3}$ | 3 | 18 | 0.19 | 0.71 | 0.11 | 0.4 | 1.3 | 2.1 | 3.0 |
| 4-P systems | 17 | 0.54 | 0.03 | 1.8 | 0.13 | $2\cdot10^{-4}$ | 0 | 21 | 0.55 | 0.07 | 0.38 | 0.7 | 1.1 | 2.0 | 3.0 |
| **Mercury** [a] | - | **0.39** | **0.21** | **6.8** | **0.055** | ? | ≥1? | ? | - | - | - | - | - | - | - |

**Notes.** N represents the number of identified representative Mercury analogs. Additionally, $a$ is the semimajor axis; $e$ is the eccentricity; $i$ is the inclination; $m$ is the planet mass; WMF is the water mass fraction; nGI is the number of giant impacts (defined by a collision of an embryo/planetesimal that is at least 10% as massive as the target body); tGI is the time of the last giant impact suffered by the planet; and $f_0$, $f_{Emb}$, and $f_{Obj}$ are the mass fractions in the form of the initial embryo, aggregated embryos, and aggregated planetesimals, respectively. $a_{0\_mass}$ indicates the source region of 90% of the planet's mass and $a_{0\_H2O}$ indicates the main source region of 90% of the water delivered to the final planet by the accretion of embryos/planetesimals. All $f_0$, $f_{Emb}$, and $f_{Obj}$ quantities are given by mean values, whereas the other quantities (except for N) represent median values.

[a] The WMF of Mercury is unknown. If Mercury acquired its large iron core via one or more giant impacts, that would suggest that nGI should be at least equal to 1.

[b] Similar to O'Brien et al. (2014), if we assume that all objects in the disk outer region carried 10% of water by weight, the WMF of the Mercury analogs would increase to $3\cdot10^{-3}$ in truncated disks.

Table 5. Representative Venus analogs obtained in three-/four-terrestrial-planet analog systems

| | N | $a$ (au) | $e$ | $i$ (deg) | $m$ ($M_\oplus$) | WMF | nGI | tGI (Myr) | $f_0$ | $f_{Emb}$ | $f_{Obj}$ | $a_{0\_mass}$ (au) | | $a_{0\_H2O}$ (au) | |
|---|---|---|---|---|---|---|---|---|---|---|---|---|---|---|---|
| all | 194 | 0.66 | 0.03 | 1.4 | 0.80 | $9\cdot10^{-5}$ | 5 | 7 | 0.10 | 0.60 | 0.30 | 0.7 | 1.2 | 1.8 | 3.1 |
| fiducial | 9 | 0.57 | 0.03 | 2.0 | 0.90 | $3\cdot10^{-3}$ | 8 | 26 | 0.02 | 0.52 | 0.46 | 0.4 | 1.8 | 2.0 | 4.0 |
| truncated | 58 | 0.65 | 0.01 | 0.6 | 0.75 | $9\cdot10^{-5}$ [b] | 5 | 6 | 0.11 | 0.65 | 0.25 | 0.7 | 1.0 | 2.0 | 3.1 |
| depleted | 61 | 0.65 | 0.03 | 1.6 | 0.74 | $5\cdot10^{-5}$ | 5 | 6 | 0.09 | 0.57 | 0.34 | 0.7 | 1.1 | 1.5 | 3.0 |
| depleted-IB | 49 | 0.67 | 0.04 | 1.7 | 0.84 | $6\cdot10^{-5}$ | 5 | 8 | 0.08 | 0.56 | 0.36 | 0.7 | 1.2 | 1.6 | 3.2 |
| peaked | 17 | 0.66 | 0.02 | 1.6 | 0.88 | $2\cdot10^{-4}$ | 3 | 3 | 0.28 | 0.68 | 0.04 | 0.4 | 1.3 | 1.9 | 3.0 |
| 4-P systems | 17 | 0.70 | 0.01 | 0.6 | 0.84 | $9\cdot10^{-5}$ | 6 | 6 | 0.09 | 0.64 | 0.26 | 0.7 | 1.0 | 2.0 | 3.1 |
| **Venus** [a] | - | **0.72** | **0.03** | **2.2** | **0.82** | **$5\text{-}50\cdot10^{-5}$** | **0-2 ?** | **?** | - | - | - | - | - | - | - |

**Notes.** N represents the number of identified representative Venus analogs. All variables are the same as defined in the caption of Table 4.

[a] The WMF of Venus is an approximate estimation (see footnote 4 in the main text). If the absence of satellites around Venus were caused by fortuitous giant impacts, that would suggest that nGI should be at least equal to 2 (Alemi & Stevenson 2006). However, the lack of an internally generated magnetic dynamo may be better explained if Venus did not experience Moon-forming-class giant impacts (Jacobson et al. 2017).

[b] Similar to O'Brien et al. (2014), if we assume that all objects in the disk outer region carried 10% of water by weight, the WMF of the Venus analogs would increase to $2\cdot10^{-3}$ in truncated disks.

**Table 6.** Representative Earth analogs obtained in three-/four-terrestrial-planet analog systems

| | N | $a$ (au) | $e$ | $i$ (deg) | $m$ ($M_\oplus$) | WMF | nGI | tGI (Myr) | $f_0$ | $f_{Emb}$ | $f_{Obj}$ | $a_{0\_mass}$ (au) | | $a_{0\_H2O}$ (au) | |
|---|---|---|---|---|---|---|---|---|---|---|---|---|---|---|---|
| all | 194 | 1.01 | 0.02 | 1.1 | 0.84 | $9\cdot10^{-5}$ | 5 | 12 | 0.11 | 0.66 | 0.23 | 0.7 | 1.2 | 1.9 | 3.1 |
| fiducial | 9 | 1.08 | 0.03 | 2.0 | 1.10 | $2\cdot10^{-4}$ | 9 | 36 | 0.03 | 0.64 | 0.32 | 0.6 | 1.9 | 1.6 | 2.9 |
| truncated | 58 | 0.98 | 0.01 | 0.6 | 0.73 | $9\cdot10^{-5}$ [b] | 5 | 8 | 0.11 | 0.70 | 0.19 | 0.7 | 1.2 | 2.0 | 3.1 |
| depleted | 61 | 0.99 | 0.03 | 1.6 | 0.84 | $7\cdot10^{-5}$ | 5 | 13 | 0.09 | 0.64 | 0.27 | 0.7 | 1.2 | 1.6 | 3.0 |
| depleted-IB | 49 | 1.08 | 0.04 | 1.8 | 0.83 | $8\cdot10^{-5}$ | 5 | 22 | 0.09 | 0.64 | 0.28 | 0.7 | 1.2 | 1.7 | 3.1 |
| peaked | 17 | 1.07 | 0.02 | 0.9 | 1.09 | $2\cdot10^{-4}$ | 3 | 6 | 0.29 | 0.67 | 0.04 | 0.7 | 1.3 | 1.8 | 2.9 |
| 4-P systems | 17 | 1.04 | 0.01 | 0.5 | 0.78 | $8\cdot10^{-5}$ | 5 | 5 | 0.11 | 0.69 | 0.20 | 0.7 | 1.2 | 2.0 | 3.1 |
| **Earth** [a] | - | **1.00** | **0.03** | **2.0** | **1.00** | $\mathbf{2.5\cdot10^{-4}}$–$\mathbf{1\cdot10^{-2}}$ | **≥1** | **20-140** | - | - | - | - | - | - | - |

**Notes.** N represents the number of identified representative Earth analogs. All variables are the same as defined in the caption of Table 4.

[a] One ocean of water sets Earth's minimum WMF. The formation of the Moon after a giant collision suggests that Earth suffered at least one such an impact, whose timing should be within the 20–140 Myr time interval.

[b] Similar to O'Brien et al. (2014), if we assume that all objects in the disk outer region carried 10% of water by weight, the WMF of the Earth analogs would increase to $2\cdot10^{-3}$ in truncated disks, which is in agreement with the results of those authors.

**Table 7.** Representative Mars analogs obtained in three-/four-terrestrial-planet analog systems

| | N | $a$ (au) | $e$ | $i$ (deg) | $m$ ($M_\oplus$) | WMF | nGI | tGI (Myr) | $f_0$ | $f_{Emb}$ | $f_{Obj}$ | $a_{0\_mass}$ (au) | | $a_{0\_H2O}$ (au) | |
|---|---|---|---|---|---|---|---|---|---|---|---|---|---|---|---|
| all | 173 | 1.46 | 0.04 | 2.5 | 0.20 | $1 \cdot 10^{-4}$ | 1 | 11 | 0.42 | 0.37 | 0.21 | 0.7 | 1.3 | 1.9 | 2.9 |
| fiducial | 5 | 1.68 | 0.13 | 9.8 | 0.12 | $3 \cdot 10^{-4}$ | 1 | 15 | 0.36 | 0.35 | 0.29 | 0.8 | 1.4 | 1.6 | 2.7 |
| truncated | 53 | 1.33 | 0.02 | 1.1 | 0.20 | $1 \cdot 10^{-4\,b}$ | 1 | 4 | 0.49 | 0.34 | 0.17 | 0.7 | 1.0 | 2.0 | 3.1 |
| depleted | 56 | 1.44 | 0.05 | 2.6 | 0.21 | $9 \cdot 10^{-5}$ | 2 | 10 | 0.36 | 0.41 | 0.23 | 0.7 | 1.6 | 1.6 | 2.9 |
| depleted-IB | 45 | 1.55 | 0.06 | 3.2 | 0.21 | $1 \cdot 10^{-4}$ | 1 | 27 | 0.39 | 0.36 | 0.24 | 0.7 | 1.6 | 1.7 | 2.9 |
| peaked | 14 | 1.53 | 0.02 | 1.6 | 0.20 | $5 \cdot 10^{-4}$ | 2 | 32 | 0.55 | 0.36 | 0.09 | 0.8 | 1.7 | 1.8 | 2.9 |
| 4-P systems | 17 | 1.41 | 0.02 | 1.5 | 0.20 | $2 \cdot 10^{-4}$ | 1 | 6 | 0.48 | 0.34 | 0.19 | 0.7 | 1.2 | 2.0 | 3.1 |
| **Mars**[a] | - | **1.52** | **0.07** | **4.0** | **0.11** | $\mathbf{1\text{-}10 \cdot 10^{-4}}$ | **1?** | **?** | - | - | - | - | - | - | - |

**Notes.** N represents the number of identified representative Mars analogs. All variables are the same as defined in the caption of Table 4.

[a] The WMF of Mars is an approximate estimation (Section 1.1). The formation of the Martian surface dichotomy suggests that the planet may have suffered one giant impact (Marinova et al. 2008).

[b] Similar to O'Brien et al. (2014), if we assume that all objects in the disk outer region carried 10% of water by weight, the WMF of the Mars analogs would increase to $3 \cdot 10^{-3}$ in truncated disks.

# APPENDIX

**Table A1.** Summary of key variables for the 17 four-planet analog systems (4-P systems)

| System # | Disk | NaM | AMD | RMC | OS | Lvf (%) | MVr | VEr | MEr | WMF | nGI | tGI (Myr) | C1 | C2 | C3 | C4 | C5 | C6 | C7 | C8 |
|---|---|---|---|---|---|---|---|---|---|---|---|---|---|---|---|---|---|---|---|---|
| 1 | T7-10_r1 | 1 | 0.0002 | 83.6 | 15.7 | 26 | 0.32 | 0.56 | 0.30 | $7 \cdot 10^{-5}$ [c] | 8 | 4 | O | O | X | O | X | X | X | O |
| 2 | T7-10_r4#1 | 0 | 0.0002 | 76.4 | 29.7 | 6 | 0.10 | 2.16 | 0.17 | $7 \cdot 10^{-5}$ [c] | 5 | 9 | O | O | O | O | X | X | X | O |
| 3 | T7-10_r4#2 | 1 | 0.0001 | 88.7 | 14.6 | 11 | 0.19 | 0.73 | 0.32 | $8 \cdot 10^{-5}$ [c] | 5 | 3 | O | O | X | O | X | X | X | O |
| 4 | T7-10_r8 | 0 | 0.0017 | 70.7 | 34.9 | 5 | 0.10 | 1.64 | 0.15 | $9 \cdot 10^{-5}$ [c] | 5 | 2 | O | O | O | O | X | X | X | O |
| 5 | T7-12_r1 | 0 | 0.0002 | 80.5 | 13.3 | 14 | 0.27 | 0.65 | 0.28 | $8 \cdot 10^{-5}$ [c] | 4 | 14 | O | O | X | O | X | X | X | O |
| 6 | T7-12_r1#2 | 2 | 0.0002 | 75.0 | 14.4 | 31 | 0.19 | 1.76 | 0.41 | $1 \cdot 10^{-4}$ [c] | 4 | 4 | O | O | X | O | X | X | X | O |
| 7 | T7-12_r4#1 | 1 | 0.0001 | 54.5 | 17.5 | 5 | 0.18 | 0.94 | 0.26 | $9 \cdot 10^{-5}$ [c] | 7 | 22 | O | O | X | O | O | X | X | O |
| 8 | T7-12_r4#2 | 0 | 0.0001 | 68.4 | 15.1 | 6 | 0.15 | 1.08 | 0.25 | $1 \cdot 10^{-4}$ [c] | 5 | 12 | O | O | X | O | X | X | X | O |
| 9 | T7-12_r4#3 | 1 | 0.0001 | 74.9 | 15.2 | 11 | 0.24 | 0.60 | 0.21 | $8 \cdot 10^{-5}$ [c] | 6 | 5 | O | O | X | O | X | X | X | O |
| 10 | T7-12_r4#4 | 0 [a] | 0.0001 | 68.3 | 17.3 | 6 | 0.12 | 1.72 | 0.16 | $1 \cdot 10^{-4}$ [c] | 4 | 12 | O | O | X | O | X | X | X | O |
| 11 | T7-12_r8 | 1 | 0.0014 | 50.7 | 18.7 | 1 | 0.18 | 0.58 | 0.09 | $8 \cdot 10^{-5}$ [c] | 8 | 53 | O | O | X | O | O | O | X | O |
| 12 | D7-10_r1_23me | 1 | 0.0003 | 54.5 | 17.8 | 34 | 0.19 | 1.11 | 0.21 | $1 \cdot 10^{-4}$ | 6 | 1 | O | O | X | O [b] | X | X | X | O |
| 13 | D7-10_r4_23me | 1 | 0.0018 | 46.9 | 18.0 | 16 | 0.22 | 2.23 | 0.47 | $9 \cdot 10^{-5}$ | 4 | <1 | O | O | X | O [b] | X | X | X | O |
| 14 | D7-10_r4_12he | 0(1) | 0.0005 | 52.7 | 20.1 | 4 | 0.10 | 2.68 | 0.18 | $3 \cdot 10^{-5}$ | 4 | 38 | O | O | O | O [b] | O | X | X | X |
| 15 | D7-12IB_r1_nc | 0 | 0.0052 | 55.5 | 34.9 | 10 | 0.17 | 1.36 | 0.14 | $2 \cdot 10^{-5}$ | 6 | 15 | X | O | O | O | X | X | X | O |
| 16 | D7-12IB_r4_nc | 0 | 0.0014 | 72.5 | 28.7 | 11 | 0.11 | 1.08 | 0.29 | $2 \cdot 10^{-5}$ | 4 | 2 | O | O | O | O | X | X | X | O |
| 17 | P9-35ir4_r19 | 0 [a] | 0.0015 | 34.2 | 18.7 | 6 | 0.10 | 0.52 | 0.22 | $9 \cdot 10^{-4}$ | 5 | 3 | O | X | X | O [b] | X | X | O | O |
| Solar system | - | - | 0.0018 | 89.7 | 37.7 | 0.1-2 | 0.07 | 0.82 | 0.11 | $2.5 \cdot 10^{-4}$ – $1 \cdot 10^{-2}$ | ≥1 | 20-140 | - | - | - | - | - | - | - | - |

**Notes.** NaM is the number of additional Mars (Mercury) analogs apart from their representative analogs, WMF is the water mass fraction, nGI is the number of giant impacts, and tGI is the time of the last giant impact of the Earth analog. The other variables are defined in caption of Table 3. "O" indicates that a system satisfied a specific constraint among C1–C8, whereas "X" indicates that it did not.

[a] Systems with planet-like objects.

[b] Systems with asteroid belts similar in orbital structure compared to that observed in the inner solar system.

[c] Similar to O'Brien et al. (2014), if we assume that all objects in the disk outer region in truncated disks carried 10% of water by weight, the WMF of the Earth analogs in systems #1–11 would increase to approximately the same value of $2 \cdot 10^{-3}$.

8